\documentclass[sigconf,nonacm,screen]{acmart}
\AtBeginDocument{%
  }

\usepackage{
  float,
  epsfig,
  wrapfig,
  graphicx,
  subcaption
}
\usepackage{colortbl}
\usepackage{enumitem}
\usepackage{graphics}
\usepackage{xparse}
\usepackage{bbding}
\usepackage{pifont}    
\usepackage{xspace}
\usepackage{wasysym}
\usepackage{multirow}
\usepackage{diagbox}    
\usepackage{booktabs}
\usepackage{makecell}
\usepackage{listings}
\usepackage{hyperref}

\AtEndPreamble{
	\hypersetup{
		colorlinks = true,
		linkcolor = purple,
		anchorcolor = purple,
		citecolor = purple,
		filecolor = purple,
		urlcolor = black
	}
}

\def\eg{\emph{e.g.}\xspace}
\def\ie{\emph{i.e.}\xspace}

\def\etal{\emph{et al.}\xspace}

\newcommand{\framework}{\textsc{WalletProbe}}
\newcommand{\walletnum}{34}

\newcommand{\wallet}{wallets}
\newcommand{\tool}{transaction security checkers}
\newcommand{\sx}{Seed Constructor}
\newcommand{\sy}{Mutator}
\newcommand{\sz}{Verifier}

\begin{document}

\title{{\framework}: A Testing Framework for Browser-based Cryptocurrency Wallet Extensions}

\author{Xiaohui Hu}
\affiliation{%
  \institution{Huazhong University of Science and Technology}
  \country{China}
  }
\email{xiaohui_hu@hust.edu.cn}

\author{Ningyu He}
\affiliation{%
  \institution{Hong Kong Polytechnic University}
  \country{Hong Kong SAR, China}
  }
\email{ningyu.he@polyu.edu.hk}

\author{Haoyu Wang}
\affiliation{%
  \institution{Huazhong University of Science and Technology}
  \country{China}
  }
\email{haoyuwang@hust.edu.cn}

\begin{abstract}
Serving as the first touch point for users to the cryptocurrency world, cryptocurrency wallets allow users to manage, receive, and transmit digital assets on blockchain networks and interact with emerging decentralized finance (DeFi) applications. Unfortunately, cryptocurrency wallets have always been the prime targets for attackers, and incidents of wallet breaches have been reported from time to time. Although some recent studies have characterized the vulnerabilities and scams related to wallets, they have generally been characterized in coarse granularity, overlooking potential risks inherent in detailed designs of cryptocurrency wallets, especially from perspectives including user interaction and advanced features. To fill the void, in this paper, we present a fine-grained security analysis on browser-based cryptocurrency wallets.
To pinpoint security issues of components in wallets, we design {\framework}, a mutation-based testing framework based on visual-level oracles. 
We have identified 13 attack vectors that can be abused by attackers to exploit cryptocurrency wallets and exposed 21 concrete attack strategies. By applying {\framework} on 39 widely-adopted browser-based wallet extensions, we astonishingly figure out \textit{all of them can be abused to steal crypto assets from innocent users}. Identified potential attack vectors were reported to wallet developers timely and 26 issues have been patched already.
It is, hence, urgent for our community to take action to mitigate threats related to cryptocurrency wallets. 
We promise to release all code and data to promote the development of the community.
\end{abstract}



\maketitle

\section{Introduction}
\label{sec:intro}

The cryptocurrency ecosystem is experiencing rapidly expansion. As of April 2025, there are 17,143 cryptocurrencies in circulation~\cite{introTokenNumber}, and the global cryptocurrency adoption rate has reached 60.61\%, with over 560 million users worldwide~\cite{introGlobalAdoption}. This surge in adoption has fueled the growth of supporting services, particularly browser-based tools such as \textit{{\wallet}} and \textit{{\tool}}. Wallets (\eg, MetaMask~\cite{metamask-intro}) enable users to manage and interact with their on-chain assets, while {\tool} (\eg, Wallet Guard~\cite{wallet-guard}) enhance asset security by linking to user wallets, analyzing submitted transactions, and issuing timely alerts.
We regard these {\wallet} and {\tool} as \textit{wallet extensions}.

\textbf{Wallet extensions deliver a spectrum of services.}
Most are multi-chain, supporting the signing of both transactions and messages~\cite{sign-tx-and-msg}. Some also provide transaction simulation, allowing users to preview transaction outcomes and estimate balance changes. Given the substantial losses caused by users’ unawareness of transaction risks, wallet extensions may integrate security alert modules to detect malicious intents embedded in signings and issue timely warnings. While {\tool} focus on identifying security threats, {\wallet} emphasize user-friendly features, enabling users to interact through intuitive UI elements, such as token searches and simple transactions (\eg, token transfers, swaps, and cross-chain exchanges).

\textbf{However, wallet extensions remain a hotspot for fraud.}
In February 2025 alone, cryptocurrency wallets were linked to losses exceeding \$1.53 billion~\cite{2025-2-wallet-losses}. The following month, a hacker compromised over 600 wallets and stole more than 930 million ARB tokens~\cite{2025-3-wallet-losses}. Wallet extensions are also vulnerable to user mistakes~\cite{cryptohacks2024}; by June 2024, scams on Trust Wallet (including ice phishing and fake giveaways) had increased dramatically~\cite{trustwalletscam}. Among existing studies, only the work by Ye~\etal specifically focuses on scam patterns that exploit wallet functionalities~\cite{ye2024interface}.
Yet beyond these, we have identified many real-world scams that exploit wallet extension features in novel ways.

\textbf{Previous research has not explored abuse related to wallet extensions.}
While previous studies have examined cryptocurrency scams~\cite{phillips2020tracing, li2023double, xia2020don, xia2024walletradar, ji2020deposafe} and summarized wallet security issues~\cite{9315193, 10.1007/978-981-16-3346-1_63, houy2023security}, they focus largely on code-level vulnerabilities from a software analysis perspective, overlooking threats arising during wallet operation. Moreover, although some attack vectors have been discussed, many wallet extensions remain exposed and new strategies continue to emerge. This underscores the need for systematic testing of wallet extensions to uncover exploitable vectors that can mislead users and result in asset losses, ultimately advancing wallet security in the cryptocurrency ecosystem.

\textbf{This work.}
To the best of our knowledge, we take the first step to explore \textit{attack vectors} within the scope of cryptocurrency wallet extensions.
More specifically, first, we summarize the general workflow of wallet extensions and abstract them into a three-layer model (\S\ref{sec:background:extension}).
Then, we propose {\framework}, an effective and automated testing framework to pinpoint attack vectors that could be visually misleading to users when interacting with wallet extensions (\S\ref{sec:framework}).
We apply {\framework} to 39 most popular browser-based wallet extensions, and uncover 13 attack vectors, which target three different modules of wallet extensions (\S\ref{sec: exploitable wallet simulator} -- \S\ref{sec: abusing wallet features}). 
For each identified attack vector, not only we detail the attack strategies, but we also try our best to quantify related financial losses and provide mitigation. 
Not that, once an attack vector is marked by {\framework}, we immediately contact the development team, providing details, as well as concrete exploits and remediation strategies.

Our major contributions can be summarized as follows.

\begin{itemize}
    \item \textbf{Framework.} To the best of our knowledge, we are the pioneers in building a comprehensive testing framework for browser-based cryptocurrency wallet extensions. 
    
    \item \textbf{Findings.} Taking advantage of {\framework}, we have uncovered 13 unique attack vectors, where six are newly discovered by us, and five were disclosed in the community while we figured out new attack strategies. 

    \item \textbf{Quantification.} We build large-scale datasets containing existing known malicious behaviors, like risky ENS names and phishing addresses. Based on them, we track fund flows and quantify financial losses for attack vectors identified.
    
    \item \textbf{Feedback.} We have disclosed attack vectors to stakeholders in affected extensions. By the time of writing, 16 of them had confirmed our findings and 26 issues had been patched with our aid. Moreover, the national vulnerability database has assigned eight vulnerability IDs.
    
\end{itemize}

\section{Background}
\label{sec:relwork}

\subsection{Ethereum Improvement Proposal (EIP)}
An Ethereum Improvement Proposal (EIP) is a formal document proposing changes or new features for Ethereum blockchains~\cite{eip-wiki}. Some EIPs even provide data structures to standardize the process of signing complex meaningful messages, such as EIP-191~\cite{eip191}, EIP-712~\cite{eip712}, and EIP-4361~\cite{eip4361}. 
Note that EIP-712 and EIP-4361 require specific fields, which are illustrated in the following listing. The \texttt{chainId} specifies the blockchain network for the following transaction execution, and the \texttt{verifyContract} refers to the contract that will verify the signature.
Different message formats often require different signing methods, as Table~\ref{tab:relationship between data formats and signing methods} shows. For instance, EIP-712 should be signed by the \texttt{eth\_signTypedData\_v4} method.

\begin{lstlisting}[morekeywords={name, version, chainId, verifyContract, salt, EIP712Domain, primaryType, domain, message}, label={lst:eip712 data structure}, linewidth=0.9\linewidth, xleftmargin=16pt]
properties: {
  types: { EIP712Domain:
           { name: string, version: string, salt: bytes32,
             chainId: uint256, verifyContract: address},
  primaryType: string,
  domain:  object,
  message: object }
\end{lstlisting}

\subsection{Transactions \& Messages}
\label{subsec:background-tx and message}
A \textit{transaction} serves as a mechanism for transferring data and value between participants, typically including essential details such as sender addresses and transferred amounts. However, signing complex \textit{transactions} on-chain often incurs significant gas fees, which can hinder efficiency and scalability. To address this issue, many decentralized applications (DApps), such as Uniswap~\cite{uniswap-permit2}, have adopted a model where users sign messages off-chain and submit only the signing hash on-chain.
These off-chain messages are typically structured according to EIPs to enhance readability and help users understand what they are about to approve.
This approach not only reduces on-chain computational costs but also enhances user experience by providing clarity and security in the signing process.
Each \textit{transaction} or \textit{message} should be submitted with its corresponding signing method, which we show in Table~\ref{tab:relationship between data formats and signing methods}.

\begin{table}[t]
    \centering
    \caption{Relationships between message data formats and signing methods.}
    \resizebox{0.8\columnwidth}{!}{
    \begin{tabular}{cc}
    \toprule
      \textbf{Format of Message Data} & \textbf{Signing Method} \\ \midrule
      hash string & \texttt{eth\_sign} \\
      text string &  \texttt{personal\_sign} \\ \midrule
      EIP-191 & \texttt{personal\_sign} / \texttt{eth\_sign} \\
      EIP-712 & \texttt{eth\_signTypedData\_v4} \\
      EIP-4361 & \texttt{eth\_signTypedData\_v4} \\ \midrule
      transaction & \texttt{eth\_sendTransactions} \\
      \bottomrule
    \end{tabular}
    }
    \label{tab:relationship between data formats and signing methods}
\end{table}

\subsection{Cryptocurrency Wallet Extensions}
\label{sec:background:extension}
Software-based cryptocurrency wallets operate across operating systems and web browsers~\cite{9315193}, enabling users to manage assets and interact with on-chain data. Complementing these, \textit{transaction security checkers} are browser extensions that intercept transactions and messages before they reach the wallet, analyze their security, and forward them to the wallet only if deemed safe.
In this paper, \textit{both cryptocurrency {\wallet} (e.g., MetaMask~\cite{metamask-intro}) and {\tool} (e.g., Wallet Guard~\cite{wallet-guard}) are considered within our scope and we refer to them as \textit{\textbf{wallet extensions}}. }

\subsubsection{External Requests for Wallet Extensions}
\label{subsubsec:external requests for wallet extensions}
We use Figure~\ref{fig:wallet components} to illustrate that external requests to wallet extensions, as \ding{182} shows, primarily involve three types of inputs: \textbf{transactions}, \textbf{messages} and \textbf{user interactions}. The first two are generated by decentralized applications (DApps) when users connect their wallet to a DApp website. These files act as an interface, allowing the DApp to submit raw data and specify the signing method required for user approval.
As for user interactions, they are initiated by users through actions such as typing text or clicking buttons within the wallet’s user interface (UI). These interactions can be formatted in JSON and consist of a sequence of actions.

\subsubsection{Components of Wallet Extensions}
\label{subsubsec:components of wallet extensions}
\begin{figure}[t]
    \centering
    \includegraphics[width=0.9\linewidth]{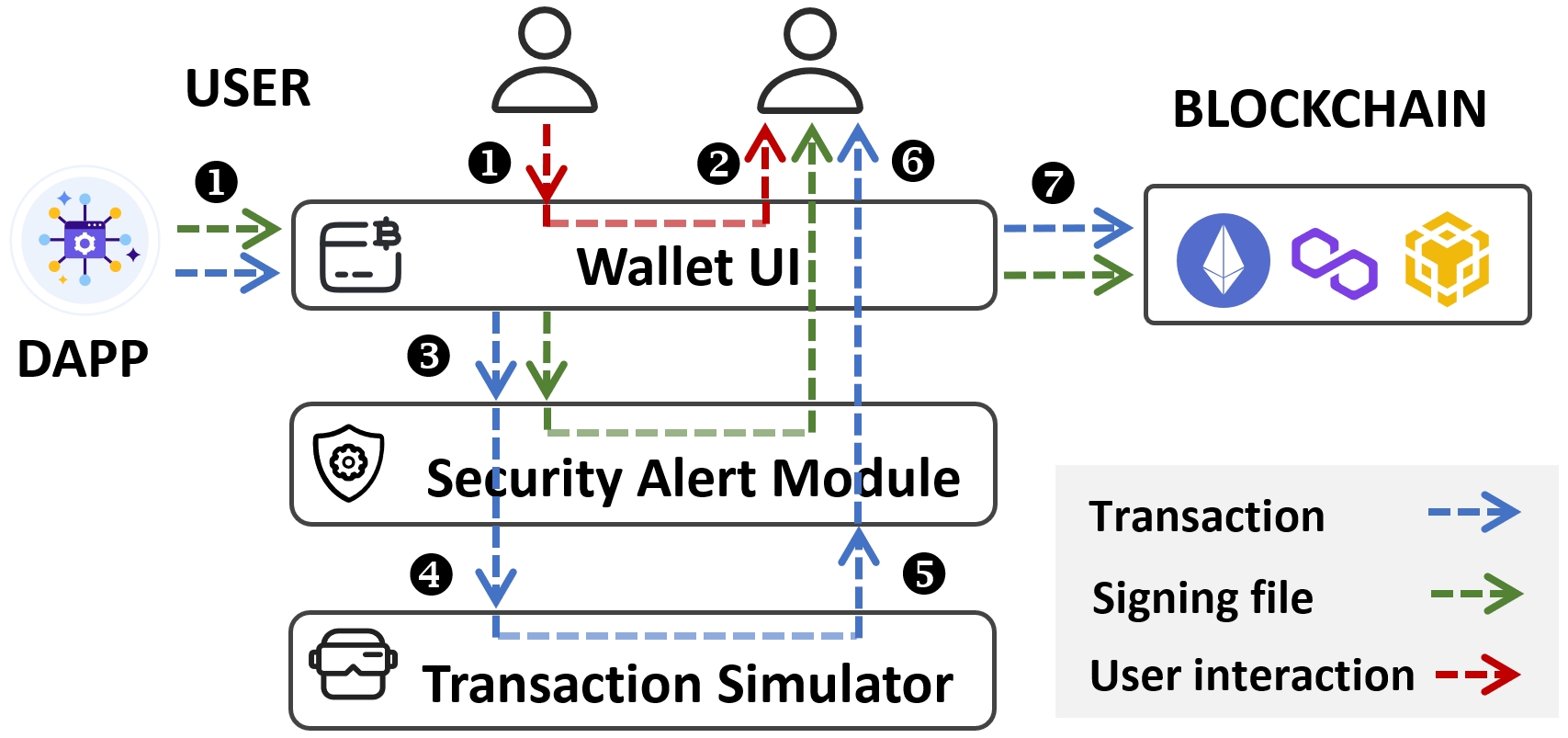}
    \caption{The workflow and architecture of wallet extensions within the Ethereum ecosystem.}
    \label{fig:wallet components}
\end{figure}

Figure~\ref{fig:wallet components} also illustrates the architecture of wallet extensions, consisting of three core components and seven crucial steps. Specifically, the \textbf{user interface (UI)} serves as the entry point for external requests, providing users with clear, visually accessible information. Beyond its foundational role, the UI often incorporates features to enhance usability, such as auto-complete and smart associations.
The \textbf{transaction simulator} processes transactions as inputs, simulating and presenting state changes for the entities involved before transactions are broadcasted on-chain~\cite{bitquery-txsimulation}. This simulation is particularly critical for helping users comprehend transaction intents, especially when complex business logic is involved. However, simulations are inherently limited in accuracy due to possible dependencies on dynamic and nondeterministic on-chain state variables (\eg, timestamp, gas limit) or user-controlled parameters (\eg, gas price). Transactions influenced by such factors are referred to as \textit{state-dependent transactions.}
At last, the \textbf{security alert module} integrates inputs from both the UI (transactions and messages) and the transaction simulator (simulated results). By analyzing the raw data and signing methods, or inferring intent from simulation outcomes, this module identifies potential risks and issues alerts to users, enhancing the overall security of the wallet ecosystem.

More specifically, the workflow of wallet extensions can be structured in a multi-phase pipeline:
\ding{182} Transactions, messages, and user interactions are submitted through the UI.
\ding{183} UI processes user inputs and displays appropriate responses.
\ding{184} Transactions and messages are forwarded to the security alert module for preliminary checks based on raw data and signing methods. If a wallet extension does not have a transaction simulator or detects potential risks, the relevant data and warnings will be presented to the user (proceeding to \ding{187}).
\ding{185} Passed transactions are delivered to the transaction simulator, which computes state changes in real time.
\ding{186} The security alert module analyzes simulation results, detects risks such as phishing transactions, and forwards findings to the UI.
\ding{187} Security check results are displayed, providing users with clear warnings and transaction details to ensure informed decision making.
\ding{188} Once signed, the transaction is immediately submitted and broadcasted to the designated blockchain. While the signed message is lately submitted after it is packed and encoded.

\section{Research Focus}
\label{sec:wallet service components}
This section presents the adversary model and scope of this work, as well as the challenges and effective solutions.

\subsection{Adversary Model \& Research Scope}
\label{subsec:adversary model and research scope}
In this work, we assume that both wallet extensions and on-chain data are accessible to both regular users and attackers. Since wallet extensions are typically invoked by users or DApp websites, attackers can also deploy malicious DApp websites, lure users into visiting them, and submit external requests to ask for signing.

Our focus is on security issues related to blockchain-specific components within wallet extensions, excluding concerns such as private key leakage and network attacks. Specifically, the goal is to identify \textit{attack vectors}, referring to wallet-specific features or design vulnerabilities that can be exploited to cause financial losses.

\subsection{Challenges \& Solutions}
\label{subsec:challenges}
We highlight two challenges that need to be addressed for the automated test, also presenting our solutions to each.

\noindent
\textbf{Challenge\#1: Generating user interactions against black-box wallet extensions.}
Code is the intellectual property for a company.
Thus, wallet extensions typically do not disclose their implementations and vary significantly in design, posing challenges for a scalable and effective generation of user interactions.
Specifically, wallet extensions use different UI layouts, which usually correspond to different UI elements. The differences in functionalities of each element and responses of the wallet extension to the interaction on them make it impossible to design a universal and efficient rule-based UI interaction sequence generation method.

\noindent
\textbf{Challenge\#2: Identifying anomaly behaviors on a visual level.}
Testing requires oracles.
However, determining whether a pop-up UI is visually confusing or not is challenging.
For instance, an address may be partially displayed due to limited display space, which prevents users from verifying it. For another example, when users enter an address, wallet extensions may auto-suggest associated ones in a drop-down box, leading to unexpected selection errors.
A visual-level oracle should be able to distinguish such ambiguities.

\noindent
\textbf{Our Solution:}
To tackle \textbf{Challenge\#1}, we design a \textit{UI-element-graph-based semantic inferring method}.
This method automatically traverses UI interfaces by simulating interactions, recording the dependency relationships of traversed UI elements, and constructing a graph.
By backtracing this graph at the given point, this method could infer its expected behaviors with UI labels and generate interactions that are syntactically and semantically correct.
For \textbf{Challenge\#2}, we design a \textit{OCR-based differential-testing-enabled anomaly identification method}.
This method combines Chrome's screen capture API and an OCR tool to analyze on-screen content. Three oracles are designed for three components in wallet extensions.
These oracles stand on two assumptions: \textit{i)} real-world execution results are ground truth, and \textit{ii)} execution with invalid or non-compliant fields must be alerted.
If OCR captured content violates either of these two principles, the corresponding attack vector will be identified efficiently against a specific component.

\section{{\framework}}
\label{sec:framework}

\begin{figure*}
    \centering
    \includegraphics[width=0.9\textwidth]{ 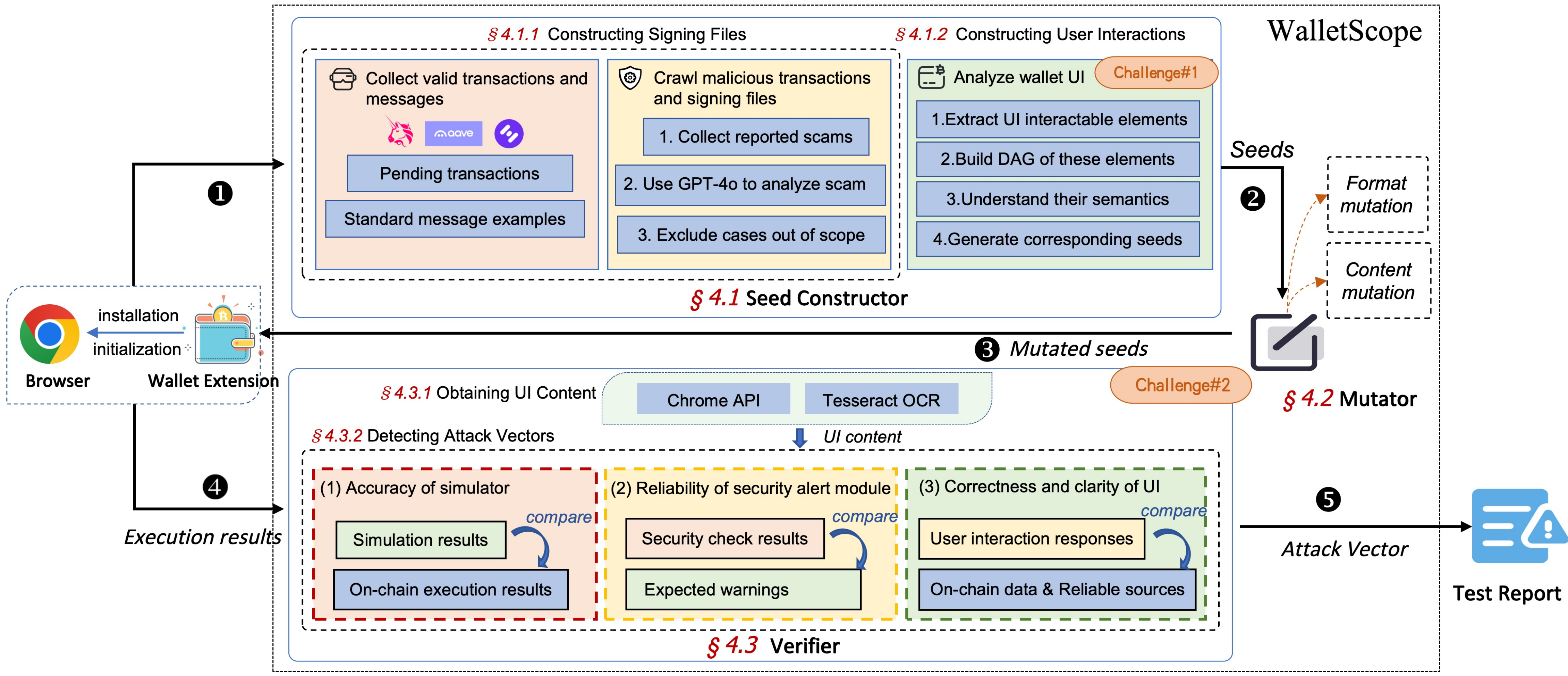}
    \caption{The workflow and architecture of {\framework}.}
    \label{fig:testing framework}
\end{figure*}

We propose a testing framework, {\framework}, whose workflow and architecture are illustrated in Figure~\ref{fig:testing framework}. The framework consists of three main components: \textit{{\sx}}, \textit{{\sy}}, and \textit{{\sz}}.
The testing process follows a pipeline: 
\ding{182} A given wallet extension will be installed and initialized in web browsers.
\ding{183} {\sx} constructs a transaction and message seed corpus, which is a one-time effort. {\sx} also automatically traverses the UI of the wallet extension to generate extension-specific UI interaction seed corpus. Both seed corpora will be passed to {\sy} for diversity.
\ding{184} {\sy} performs content and format mutation on passed seeds, which will then be passed to the wallet extension.
\ding{185} {\sz} captures the wallet UI to parse the execution results, based on which {\sz} integrates three oracles corresponding to three components in wallets to determine whether a seed can be exploited as an attack vector. 
\ding{186} Confirmed attack vectors are finally organized into the test report.
We delve into the design and implementation of {\framework}.

\subsection{{\sx}}
\label{subsec: implement seed generator}
{\sx} undertakes constructing an external request corpus (see \S\ref{subsubsec:external requests for wallet extensions}) with transactions, messages, and user interactions.

\subsubsection{Collecting Transactions and Messages}
\label{subsubsec:signing files generator}
To ensure the diversity of seeds, {\sx} seeks to construct the corpus from various data sources.

\noindent
\textbf{Collecting valid transactions and messages.}
First of all, we collect valid transactions from the mempool, and messages from EIP documentations.
Specifically, for transactions, we deploy a full node through Geth~\cite{geth} as a listener. It subscribes to newly propagated transaction hashes and retrieves the corresponding transaction's metadata upon detection. 
For valid messages, since wallet extensions allow DApps or users to submit messages structured in various formats, along with various signing methods (see Table~\ref{tab:relationship between data formats and signing methods}), it is hence intuitive to collect as many types of message formats as possible.
To this end, we develop a crawler to scan the documentation of EIPs and extract all given examples~\cite{eip191, eip712, eip4361, eth-sign-metamask, personal-sign, eth-signTypedData}.

\noindent
\textbf{Crawling malicious transactions and messages.} 
Existing malicious instances can be used directly or mutated to verify the robustness and sensitivity of wallet extensions.
Thus, we build a crawler to collect reported scam events on reliable platforms~\cite{chainabuse, slowmisthascked} to collect corresponding transactions and messages. 
Specifically, it scans reports and takes advantage of GPT-4o to analyze text, extracts message data or transaction hashes highlighted in the text. The crawler then retrieves transaction metadata from the blockchain with hashes and annotates each message or transaction with a determined type of scam. 
We then ask two involved researchers to exclude cases that are out of our research scope (see \S\ref{subsec:adversary model and research scope}), \textit{e.g.,} a social engineering scam.

\subsubsection{Constructing User Interactions}
\label{subsubsec: user interaction generator}
Wallet UIs provide various functionalities (\eg, ENS resolution, token search, and metadata editing) whose implementations are often closed-source. 
As discussed in \textbf{Challenge\#1}, how to automatically recognize and trigger them is not a trivial task.
To address this challenge, we design a \textit{UI element graph-based semantics inferring method}.

\begin{figure}
    \centering
    \includegraphics[width=\linewidth]{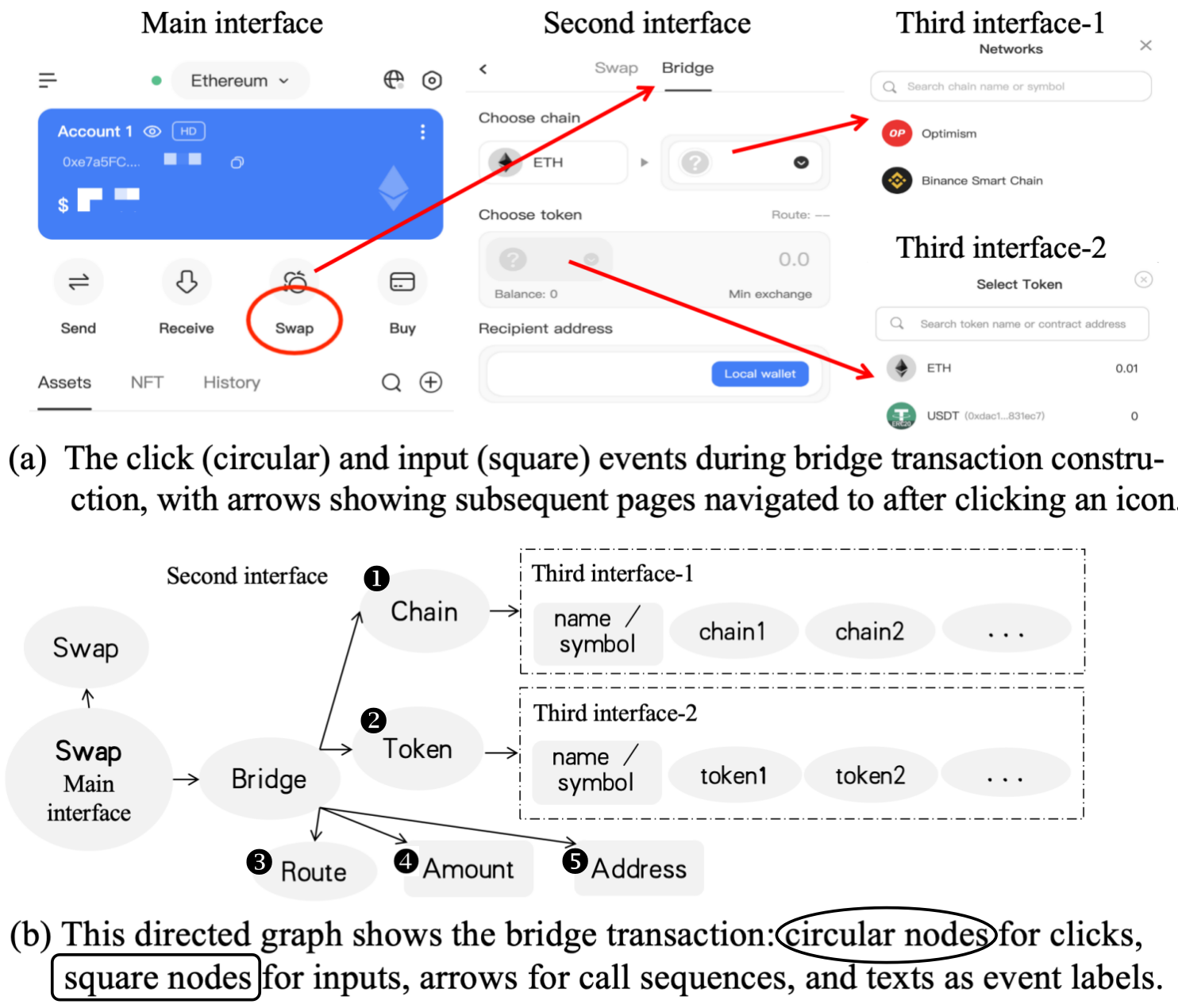}
    \caption{The user interface and its corresponding DAG.}
    \label{fig:directed figure to swap}
\end{figure}

Specifically, the method leverages Puppeteer~\cite{puppeteer}, a widely used browser automation tool, to parse the UI of a given wallet extension and extract all interactable elements, such as input boxes and buttons. It simulates user interactions (\eg, button clicks) to traverse all UI pages and build a directed acyclic graph (DAG) where nodes represent interactable UI elements and edges denote their dependency relationships.
For example, Figure~\ref{fig:directed figure to swap}{\color{purple}{(a)}} illustrates a UI flow for initiating a bridge transaction: clicking the \textbf{Swap} button opens a new UI interface where users select the target chain, token, amount, and recipient address. A third UI interface prompts users to reconfirm the chain and token.
Figure~\ref{fig:directed figure to swap}{\color{purple}{(b)}} shows the generated DAG, where circles and rectangles indicate clickable elements and input fields, respectively.
From this DAG, the method can infer the semantics of an element by \textit{backtracking its predecessors}. For instance, backtracking the \textit{Amount} node yields the path \textit{Swap → Bridge → Amount}, indicating the context is a bridge transaction and the input type is a token amount.

Leveraging the DAG and inferred element semantics, {\sx} constructs seeds for each interactable path in the DAG. 
Specifically, {\sx} starts with text input nodes, backtracks the graph, and ends at the entry. Therefore, {\sx} could determine the data format required for the text input field according to the semantics derived previously. For example, the amount field must be an integer. 
Consequently, {\sx} will generate a JSON file consisting of elements indicating the interaction type (like \textit{click} or \textit{input}), random data of the designated type (like \textit{integer} or \textit{address}), and the triggering path.

\subsection{{\sy}}
\label{subsec: sy}
To further improve the diversity of constructed seeds, {\sy} adopts \textit{semantics aware mutation strategies} on seeds. 
Although in \S\ref{subsubsec: user interaction generator}, we can obtain the semantics of UI elements, with respects to the fields of transactions and messages, we try to infer their semantics according to the name (\eg, the \texttt{value} field in transactions specifies the native currency amount transferred) and data format (\eg, a sequence of 24 zeros followed by 40 characters can be inferred as an address). 
Then, we apply a hybrid mutation, including content mutation and format mutation. We illustrate their details in Table~\ref{tab:mutation strategies}.

\begin{table*}[htbp]
  \centering
  \caption{Mutation strategies and examples.}
    \resizebox{0.85\textwidth}{!}{
    \begin{tabular}{c|l|l}
    \midrule
          & \textbf{Content Mutation} & \textbf{Format Mutation} \\
    \toprule
    \multirow{8}[2]{*}{\textbf{\makecell{Transaction\\ \& \\Message}}} & 1. altering the transfer value to zero or a very large number; &  \\
          & 2. replacing an address with another one; & \\
          & 3. randomly truncating, inserting, and replacing characters;  & 1. altering data types;  \\
          & 4. illegal encoding; & \quad(\eg, converting hexadecimal to decimal)\\
          & \quad(\eg, introducing unreadable characters) &  2. swapping the order of parameters;\\
          & 5. normal mutation strategies; &  \\
          & \quad(\textit{e.g.,} bit flipping and arithmetic operation mutations) & \\
    \midrule
    \textbf{User interaction} & above all & above all \& changing the order of UI elements in the path \\
    \toprule
    \end{tabular}%
}
  \label{tab:mutation strategies}%
\end{table*}%

\subsection{{\sz}}
After the wallet extension processes a seed, {\sz} is responsible for obtaining dynamic execution results and applying oracles to detect possible attack vectors.

\subsubsection{Obtaining UI Content}
{\sz} obtains the execution results by screen capturing, whose reasons are twofold.
On the one hand, most wallet extensions are closed-source, and {\sz} is unable to intercept execution traces through instrumentation. On the other hand, simulation results, security check results, and UI responses are all ultimately presented to users through UI. 
We have introduced in \S\ref{subsec:challenges} that conducting visual-level verifications is a non-trivial challenge in \textbf{Challenge\#2}. To this end, we propose an \textit{OCR-based differential-testing-enabled anomaly identification method}. 
To be specific, {\sz} takes advantage of a Chrome API, \texttt{captureVisibleTab()}, to capture the currently active tab in the specified window. Then it adopts Tesseract~\cite{tessertact}, an OCR engine, to recognize all text information and layout.

\subsubsection{Detecting Anomalies in Components}
For different components in wallet extensions, we design specific methods to identify anomalies and present them as follows.

\noindent
\textbf{Verifying the accuracy of the simulator.} 
Simulator takes transactions as input (see \ding{185} in Figure~\ref{fig:wallet components}), thus {\sz} determines its accuracy by comparing simulation results and actual on-chain results. 
First, some collected valid transactions (see \S\ref{subsubsec:signing files generator}) are taken as inputs. {\sz} captures the UI after the simulation completes and performs keyword matching to determine the simulation result. For example, the ``+'' or ``receive'' indicates that the user will receive certain tokens; if ``fail'' occurs on the UI, it suggests the transaction will fail on-chain.
Then, {\sz} retrieves the corresponding transaction metadata on-chain and investigate its results for comparison.
Moreover, {\sz} also expects mutated transactions with non-compliant or invalid value of any fields to fail in the simulation environment (\eg, the token address is replaced with a meaningless string).
If a transaction violates any of the above two expected behaviors, it will be considered a possible attack vector.

\noindent
\textbf{Verifying the reliability of the security alert module.}
A security alert module takes transactions and messages as input (see \ding{184} in Figure~\ref{fig:wallet components}), thus {\sz} determines its reliability with non-benign transactions or messages as inputs.
To achieve this, {\sz} tries to match keywords such as ``warning'' and ``risk'' after capturing UI. Note that, if UI remains unchanged after submitting a seed, {\sz} interprets it as an implicit alert, indicating that users are prevented from proceeding.
{\sz} also expects that \textit{i)} non-compliant value in critical fields and \textit{ii)} inconsistent environment between the signed data and the connected one should trigger alerts.
If the above expected behavior does not occur, the input transaction or message will be considered as a possible attack vector.

\noindent
\textbf{Verifying the correctness and clarity of UI.} 
The UI takes user interactions as input (see \ding{182} in Figure~\ref{fig:wallet components}), thus {\sz} determines its correctness and clarity by checking UI responses to user interactions, which can be directly obtained through captured UI content and layout. 
Regarding the correctness, {\sz} compares UI content with on- and off-chain records. For example, it expects ENS name resolutions and token metadata to be identical to the on-chain ones. It also expects the auto-suggestion of a token name can be successfully retrieved from widely recognized off-chain websites, \ie, CoinMarketCap~\cite{coinmarketcap}, CoinGecko~\cite{coingecko}, and TokenInsight~\cite{tokeninsight}.
As for the clarity, {\sz} pays attention to the presentation format, like whether metadata is displayed as raw JSON directly or in the form of key-value. {\sz} further examines whether important information is presented, \textit{i.e.,} the data that can influence the transfer of funds, like token and recipient addresses in transactions and the \texttt{chainID} field in an EIP-712 message. 
If a UI interaction seed violates either the correctness or clarity principles, it is considered an attack vector.

\section{Results Overview \& Ethical Considerations}
In this section, we will conduct a high-level overview of obtained results and underline ethical considerations.

\subsection{Experiment}

\noindent
\textbf{Experimental Setup \& Implementation.}
We deploy a client node via Geth~\cite{geth} connected to the Sepolia testnet to submit transactions, which are synced and retrieved from Ethereum mainnet via another Geth node. 
As for selecting wallet extension candidates, we go through the Chrome Webstore~\cite{chromestore} to look for the ones whose downloading time is exceeding 1,000.
Finally, 39 the most popular wallet extensions are included, accounting for 110M monthly active users.
At the first step in Figure~\ref{fig:testing framework}, all wallet extensions are downloaded from the official Chrome Store, installed and initialized with default settings.
The only manual intervention was to switch them to the Sepolia testnet for subsequent testing.
{\framework} is implemented in JavaScript and Python with 3.2K LOC. It also takes advantage of Tesseract-0.3.13 to capture wallet UI content.

\noindent
\textbf{Results Overview.}
Out of 39 candidates, there are {\walletnum} {\wallet} and five {\tool}.
Targeting them, {\framework} has raised 158 alerts in total. Based on the impacted components and concrete behaviors, 13 attack vectors are identified.
More specifically, six are newly discovered by us; five were disclosed in the community while we figure out new attack strategies; and two were explored in prior research.
{\framework} only marks six false positives (FPR=6/39=15.4\%) when identifying $V_{13}$ as the limitation that {\framework} cannot scroll the UI page to capture contents on screen.
The average time for each wallet is 22 minutes, primarily spending on constructing UI element DAG.

Astonishingly, all wallet extensions are affected by at least three attack vectors, and there are 4.9 attack vectors for each on average. 
Compared to the other two components, the security alert module is more vulnerable, corresponding to eight attack vectors.
After a comprehensive manual examination on these reports, we conclude this is because \textit{i)} a single risky transaction pattern may be implemented in diverse ways where wallet extensions always have limited detection scope, and \textit{ii)} structured messages bring challenges for wallets to effectively identify the embedded malicious intent, though they significantly enhance user experience.

After our timely disclosure to the corresponding developer teams, 16 ones have confirmed all our reported attack vectors and conveyed gratitude, and 26 issues have been patched already. 
All identified attack vectors have been confirmed by at least two teams separately, illustrating the real-world impact of our identified attack vectors.
We have also reported vulnerabilities to the national vulnerability database and, as of now, have been assigned eight IDs.

\subsection{Ethical Considerations}
First of all, we have to underline that all wallet extension candidates are publicly accessible and we only use the release versions published by official teams in the Chrome Store.
Once an attack vector is reported and confirmed by {\framework}, we report the identified attack vector to developers through email or direct messages on social media within 30 minutes. 
Specifically, for each specific attack vector, we provide the corresponding exploits\footnote{Exploits in this work vary in forms, like a transaction or a series of UI element interactions.} and the evidence of its impact, which is often a screen capture, illustrating the possible misleading of the attack vector can lead to. 
Of course, we also provide detailed attack mitigation strategies and, where applicable, references to existing work that can assist in optimizing security solutions. 
After timely disclosure, we will continue to track the progress of the fix and express our willingness to provide support and assistance to the development team in any way we can.


In the following, \S\ref{sec: exploitable wallet simulator} to \S\ref{sec: abusing wallet features} illustrate attack vectors against the transaction simulator, security alert module, and wallet UI, respectively. 
For each, we detail how to abuse, the impact quantification, interesting findings, and mitigation methods.
\section{Simulation is Not Reality}
\label{sec: exploitable wallet simulator}
In this section, we disclose two attack vectors ($V_1$ and $V_2$). Currently, ten wallet extensions support the transaction simulator, \textit{i.e.,} Coinbase, MetaMask, Rabby, TokenPocket, Zeal, AegisWeb3, Web3Antivirus, WalletGuard, Fire, and PocketUniverse. Our analysis reveals that all ten extensions are vulnerable to $V_1$, while only Rabby and MetaMask are susceptible to $V_2$.

\subsection{Exploitable Transaction Dependency*\protect\footnote{Targeting cryptocurrency wallet extensions, \# indicates the attack vector is newly disclosed by us; $\star$ indicates the attack vector has been revealed, but we disclose new attack strategies; $\bullet$ indicates the attack vector has been explored.} ($V_1$)}

\noindent
\textbf{Attacks.}
Exploitable transaction dependency ($V_1$) occurs when $tx_1$ relies on states modified by $tx_2$ earlier \textit{in the same block}.
As simulators must have no idea of the existence of $tx_2$, the simulation of $tx_1$ must be different from the on-chain executed result, where $tx_2$ will be processed at first.
{\sz} identified this attack vector by comparing the actual execution results of the crawled transaction (see \S\ref{subsubsec:signing files generator}) with its local simulation results.
Attackers can exploit $V_1$ by front-running $tx_1$ with a crafted $tx_2$, which manipulates a state variable that $tx_1$ depends on, tricking users with the locally misleading simulation results.

\noindent
\textbf{Impact.}
All ten wallet extensions supporting the transaction simulator are vulnerable to $V_1$.
This is because they can only perform simulation on a single transaction, which may open the door to \textit{front-running MEV attacks}~\cite{chi2024remeasuringarbitragesandwichattacks}, as we mentioned above.
By exploiting $V_1$, attackers can easily manipulate price movements to their advantage, leading to considerable and unexpected financial losses for users, who pay substantially more for a token or receive far less than the anticipated amount shown by the local simulator.

\noindent \textbf{Mitigation.}
Unfortunately, $V_1$ is an inherent limitation for simulators. 
To mitigate its financial impact, wallet extensions can: 
\textit{i)} maintain a blacklist of contracts that are prone to being front-ran. If a user attempts to interact with one of these contracts, the wallet will issue an explicit warning about the possibility of being front-ran;
and \textit{ii)} leverage tools like IcyChecker, a fuzzing-based framework, to detect potential front-running vulnerabilities in smart contracts~\cite{10.1145/3597926.3598057} and warn users as well.

\subsection{Manipulable State Variable$^\#$ ($V_2$)}

\begin{figure}[t]
   \centering
   \includegraphics[width=\linewidth]{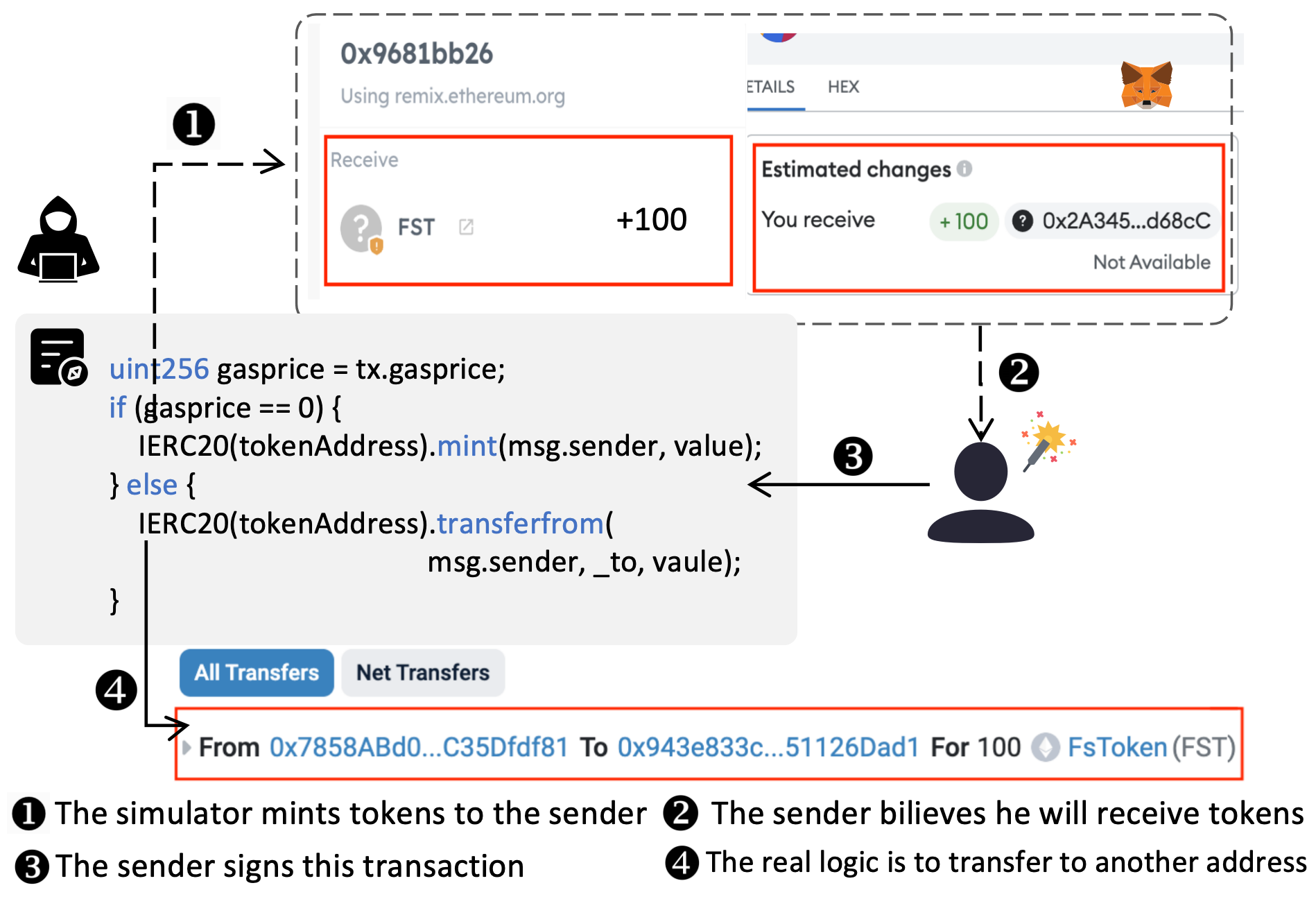}
   \caption{An example exploits $V_2$, where the user deceived will receive 100 tokens according to the simulation results.}
   \label{fig:wrong simulation}
\end{figure}

\noindent
\textbf{Attack.}
Manipulable state variable ($V_2$) exploits those state variables that the simulator cannot predict but has default values for, \eg, \texttt{tx.gasprice} is a variable while simulators often set it as zero.
Figure~\ref{fig:wrong simulation} presents a concrete example where attackers exploit $V_2$.
As we can see, the attacker embeds an if-else statement in the attack contract, where the condition depends on the value of \texttt{gasprice}. If a transaction is initiated to this contract, the contract behavior will be different depending on whether it is in a local simulation environment or a real on-chain environment. To this end, the attacker can put some logic that lures users under the if branch, and put the actual behavior that will be executed on-chain (such as transferring the user's asset) under the else branch.

\noindent
\textbf{Impact.}
This type of attack is completely imperceptible to users, especially when the contract is not open source.
To evaluate if such attack vectors are adopted in real-world, we analyze contracts deployed between January and November 2024 with static analysis. 
In total, we have identified 26 contracts that rely on manipulatable state variables, \textit{e.g.,} \texttt{block.coinbase} and \texttt{tx.gasprice}, in the condition of an if-else statement. More importantly, there are external calls or transfers under the branches, as illustrated in  Figure~\ref{fig:wrong simulation}. 
On average, each of them are involved in 107.58 transactions, accounting for \$60K financial losses.

Regarding the number of manipulatable state variables, we have identified seven in total, which can be exploited in MetaMask and Rabby\footnote{Other wallet extensions will not display simulation results for transactions involving manipulatable state variables.}. 
For example, both wallets set \texttt{tx.gasprice} to zero when simulating transactions on Ethereum mainnet and use a fixed set of addresses for \texttt{block.coinbase} on Sepolia. This will not only result in inaccurate simulations, but also lead to an attack vector that could be exploited.

\noindent \textbf{Mitigation.}
The most intuitive way is to hide the simulation results of state-dependent transactions, as the
other wallet extensions, since such inabilities are inherent limitations for the simulator. This can be achieved be integrating static analysis tools, \eg, Gigahorse~\cite{gigahorse}
After our disclosure, both Metamask and Rabby accepted our advice and adopted this strategy immediately, though it somehow sacrifices user experience.

\section{Alert Now or Never}
\label{sec: vulnerable security system}
The security alert module is supposed to detect malicious behaviors in real time, which in turn discourages users from manually verifying transaction data. Thus, if a malicious seed evades the security check, users will believe it to be safe and unwittingly sign it. 
We have identified eight attack vectors, as shown in Table~\ref{tab:security alert module result}.

\begin{table*}[htbp]
  \renewcommand{\arraystretch}{1.1}
  \centering
  \caption{The feasibility of attack vectors against the security alert module of wallets and \underline{security checkers}. Space indicates the wallet is not vulnerable, \ding{55} indicates the wallet is vulnerable, \CIRCLE{} indicates that the wallet has already fixed this vulnerability.}
  \resizebox{0.9\textwidth}{!}{
    \begin{tabular}{cccccccccc||cccccccccc}
    \toprule
     \textbf{Name} & \textbf{Confirmed} & {$V_3$} & $V_4$ & $V_5$ & $V_6$ & $V_7$ & $V_8$ & $V_9$ & $V_{10}$ & {\textbf{Name}} & {\textbf{Confirmed}} & {$V_3$} & {$V_4$} & {$V_5$} & {$V_6$} & {$V_7$} & {$V_8$} & {$V_9$} & {$V_{10}$} \\ \midrule
     \rowcolor{white} Aurox~\cite{aurox} &  \ding{51} &    & \ding{55}   & \ding{55}   &       & \ding{55}   &       & \ding{55}   &    & {Phantom}~\cite{phantom} & \ding{51} &       & {\CIRCLE{} } & {\CIRCLE{} } & {\ding{55} } &      {\CIRCLE{} } & {\ding{55} } &  & \\
     \rowcolor[rgb]{ .949,  .949,  .949}{Argent}~\cite{argentx-intro} &  &   &  &  &  &  &  & {\ding{55}} &       & {Rabby}~\cite{rabby} & \ding{51} &     &  &       & {\CIRCLE{} } &       &       & {\ding{55} } &     \\
    \rowcolor{white}{Braavos}~\cite{braavos} &  &   &  &     &     &     & {\ding{55} } & {\ding{55} } &   &{Rainbow}~\cite{alchemy-integrated-rainbow} &  \ding{51} &  
     &{\ding{55} } & {\ding{55} } & {\ding{55} } & {\ding{55} } &  &
     {\ding{55} } &    \\
    \rowcolor[rgb]{ .949,  .949,  .949} Bitget~\cite{bitget} &  \ding{51}  &   & \ding{55}   & \ding{55}   & \ding{55}   & \ding{55}   & & \ding{55}   & \CIRCLE   &  {SafePal}~\cite{safepal} &   &    \ding{55}  & {\ding{55} } & {\ding{55} } & {\ding{55} } & {\ding{55} } & {\ding{55} } & {\ding{55} } &  \\
    \rowcolor{white} Bybit~\cite{bybit} &  &  \ding{55}  &       & \ding{55}   & \ding{55}   & \ding{55}   & \ding{55}   & \ding{55}  &       &  {Sequence}~\cite{sequence} &    &  \ding{55} &       & {\ding{55} } & {\ding{55} } & {\ding{55} } & {\ding{55} } & {\ding{55} }  &  \\
    \rowcolor[rgb]{ .949,  .949,  .949} Core~\cite{core}  &   &   &       & \ding{55}   & \ding{55}   & \CIRCLE   & \CIRCLE  & \ding{55}  & \ding{55}   &  {Sender}~\cite{sender} & \ding{51} &  \ding{55} &    & {\ding{55} } & {\ding{55} } & {\ding{55} } & {\ding{55} } & {\ding{55} } &  \\
    \rowcolor{white} Coin98~\cite{coin98} &  &      & \ding{55}   & \ding{55}   & \ding{55}   & \ding{55}   &  \ding{55}  & \ding{55}   &    &  {SubWallet}~\cite{subwallet} &  \ding{51} &   \ding{55} &  & {\ding{55} } & {\ding{55} } & {\ding{55} } & {\ding{55} } & {\ding{55} } &  \\
    \rowcolor[rgb]{ .949,  .949,  .949}   Coinbase~\cite{coinbase} &  \ding{51}    &    &   & \CIRCLE{}  & \CIRCLE{} &       &       & \ding{55}  &       & {Trust Wallet}~\cite{trustwallet} &  \ding{51} & \ding{55} &      & {\ding{55} } & {\ding{55} } & {\ding{55} } &  & {\ding{55} } &    \\
    \rowcolor{white} Enkrypt~\cite{enkrypt} &  \ding{51}  &  & \ding{55}   & \ding{55}   & \ding{55}   & \ding{55}   &       & \ding{55}   &      & {TokenPocket}~\cite{token-pocket} & \ding{51} &  &{\ding{55} } & {\ding{55} } & {\ding{55} } & {\ding{55} } &       &    &  \\
    \rowcolor[rgb]{ .949,  .949,  .949}  Exdus~\cite{exdus} &  \ding{51}    &     &  & \CIRCLE{} & \ding{55}   & \ding{55}   &       & \ding{55}   &       & {Virgo}~\cite{virgo} &  &  \ding{55}  &       & {\ding{55} } & {\ding{55} } & {\ding{55} } & {\ding{55} } & {\ding{55} } &  \\
    \rowcolor{white} Frontier~\cite{frontier} &  &       & \ding{55}   & \ding{55}   & \ding{55}   & \ding{55}   & \ding{55}   & \ding{55}   &  \ding{55} & {WigWam}~\cite{wigwam} &  \ding{51} & & {\ding{55} } & {\ding{55} } & {\ding{55} } & {\ding{55} } & {\ding{55} } & {\ding{55} } &  \\
    \rowcolor[rgb]{ .949,  .949,  .949}   Gate~\cite{gate}  &  &  \ding{55}  &       & \ding{55}   & \ding{55}   & \ding{55}   & \CIRCLE  & \ding{55}   &     & {XDEFI}~\cite{xdefi} &   &  &       &       & {\ding{55} } & {\ding{55} } &       &    &  \\
    \rowcolor{white} Hana Wallet~\cite{hana} &  & \ding{55} & \ding{55}   & \ding{55}   & \ding{55}   & \ding{55}   & \ding{55}   & \ding{55}   &       & {YeTi}~\cite{yeti} &   &   \ding{55}   &       &       & {\ding{55} } & {\ding{55} } & {\ding{55} } &       &   \\
    \rowcolor[rgb]{ .949,  .949,  .949}   Klever~\cite{klever} &  &  &       & \ding{55}   & \ding{55}   & \ding{55}   &       & \ding{55}   &    & {Zeal}~\cite{zeal} &   &   &       &       & {\ding{55} } &       &    \CIRCLE   & {\ding{55} } & {\ding{55} }  \\
    \rowcolor{white} Liquidity~\cite{liquidity-wallet} &  \ding{51} &   & \ding{55}   & \ding{55}   & \ding{55}   & \ding{55}   &       & \ding{55}   & \ding{55}  & {Zerion}~\cite{zerion} &  &    \ding{55} &       &       & {\ding{55} } &       &       & {\ding{55} } &  \\
    \rowcolor[rgb]{ .949,  .949,  .949}  MetaMask~\cite{metamask-intro} &    &   &    &  \CIRCLE{}  & \ding{55}   & \ding{55} &       &  \CIRCLE{}  &       & \underline{AegisWeb3}~\cite{aegisweb3} &   &  \ding{55}   &       & \ding{55}   & \ding{55}   &      &       &       & \\
    \rowcolor{white} OKX~\cite{okx}  &  \ding{51}    &    &   & \ding{55}   & \ding{55}   &       &       & \ding{55}   &      & \underline{Fire}~\cite{fire-wallet} &       &       &       &       & {\ding{55} } &       &       &       & \\
    \rowcolor[rgb]{ .949,  .949,  .949}  OneKey~\cite{onekey} &  & & \CIRCLE{} & \CIRCLE{}  & \ding{55}   & \ding{55}   & \CIRCLE{}   & \CIRCLE{}  & \ding{55}   & \underline{WalletGuard}~\cite{wallet-guard} &  &   &       &       &       & \ding{55}   &       & \ding{55}   &   \ding{55}   \\
    \rowcolor{white} Pail~\cite{pail}  & &   \ding{55}  &       & \ding{55}   & \ding{55}   & \ding{55}   &       & \ding{55}   &   \ding{55}   & \underline{Web3Antivirus}~\cite{web3Anti} &   &  &       &       & \ding{55}   &       &       &        &\\
    \rowcolor[rgb]{ .949,  .949,  .949}   &   &  &    &       &   &   &       &     &       & \underline{PocketUniverse}~\cite{pocketuniverse} &    &   &       & {\ding{55} } &   & {\ding{55} } &       & {\ding{55} } &   \\
    \bottomrule
    \end{tabular}
}
\label{tab:security alert module result}
\end{table*}

\subsection{Dangerous \texttt{eth\_sign}$^\bullet$ ($V_3$}

\noindent
\textbf{Attacks.}
The \texttt{eth\_sign} method~\cite{eth-sign-metamask}, which signs hashed messages, enabling attackers to hide malicious intents ($V_3$), due to which it is recommended to disable by its proposer, \textit{i.e.,} Metamask~\cite{metamask-intro}. 
{\sz} detected $V_3$ when a signing request used \texttt{eth\_sign}, yet some wallets issued no warnings. As a result, users saw only an unreadable hex string, unable to verify if they were authorizing sensitive actions like token transfers or votes.

\noindent
\textbf{Impact.}
Surprisingly, 12 wallet extensions still support \texttt{eth\_sign} without warnings. Phishing scams exploiting $V_3$ have already caused over \$70 million losses~\cite{he2023txphishscope}.
We alerted the 12 affected wallets, but only three responded. As of November 2024, the Chrome Store has blocked six of the remaining nine, flagging them as unsafe. This highlights the urgent need to treat \texttt{eth\_sign} as a high-risk feature.

\noindent
\textbf{Mitigation.}
As the proposer of \texttt{eth\_sign}, MetaMask has disabled it by default and discouraged users from enabling it on any websites~\cite{eth-sign-metamask}.
Thus, the intuitive solution is to intercept and analyze the communication between the browser and the blockchain network, capture requests for \texttt{eth\_sign} signings, and raise timely alerts.

\subsection{Non-rigorous Inputdata$^\#$ ($V_4$)}
\label{subsec: non-rigorous inputdata}

\noindent
\textbf{Attacks.}
The first four bytes of a transaction's inputdata are the signature of the invoked function, based on which wallet extensions infer malicious activities.
For transactions that call malicious functions and should trigger alerts, we found that mutating the format of the input data without changing the semantics of the input data can evade detection ($V_4$).
{\sz} discovered $V_4$ when a transaction invokes a malicious function with mutated inputdata format (like omitting the 0x prefix) without triggering any alerts.
Interestingly, we also observe that the inputdata will be reformatted to the standard format upon on-chain submission by connected client nodes, preventing the attack transaction from disrupting execution. This subtle yet effective evasion technique is identified and disclosed by us for the first time.

\noindent
\textbf{Impact.}
Results indicate that 14 wallet extensions are vulnerable to $V_4$. Unfortunately, as the mutated data will be reformatted into the standard format on-chain, such attacks cannot be detected through on-chain transaction analysis.
We identified an interesting but critical vulnerability in Bitget, which always automatically switches the connected network to Ethereum mainnet when the \texttt{0x} prefix of inputdata is removed. 
This behavior, likely caused by a software bug, can lead to transaction failures and gas fee waste. 
Even more concerning, attackers could exploit $V_4$ to deceive Bitget users into believing that transactions will be executed on the testnet, while in reality, assets are being stolen on the mainnet. 
After our disclosure, Bitget promptly patched this vulnerability.

\noindent
\textbf{Mitigation.}
The mitigation of $V_4$ is intuitive. 
On the one hand, developers of wallet extensions could adopt rigorous validation rules on all data fields within transactions to prevent such format-mutation-based evasion.
On the other hand, from a user-friendly perspective, if such non-rigorous formats are acceptable, at least there should be a non-implicit reformat before the validation on all data fields, including inputdata.

\subsection{Overlooked Approval* ($V_5$)}

\noindent
\textbf{Attacks.}
Approval in blockchain allows others to spend tokens on a user’s behalf. 
As a highly sensitive operation, if the intended notification of approval behaviors is bypassed, it may provide a possible attack vector ($V_5$).
{\sz} identifies $V_5$ by showing that some wallet extensions notify users about standard \texttt{approve()} calls, but such alerts disappear when \textit{i)} using advanced approval implementations, like the ones shown in Table~\ref{tab:malicious patterns} in Appendix, or \textit{ii)} removing the \texttt{0x} prefix from the inputdata (whose reason is explained in \S\ref{subsec: non-rigorous inputdata}).
Interestingly, {\sz} also reveals a critical shortcoming for validating the approval behaviors, \textit{i.e.,} \textit{semantic insensitivity}. For example, an unlimited approval \texttt{setApprovalForAll()} will trigger an alert, but the semantic equivalent one \texttt{approve(amount=0xff)} will not.

\noindent
\textbf{Impact.}
We have identified 29 wallet extensions that are vulnerable to $V_5$.
Specifically, 26 and 14 are vulnerable to \textit{i)} advanced approval implementations and \textit{ii)} removing \texttt{0x} prefix from inputdata, respectively.
Moreover, 20 wallet extensions suffer from the semantic insensitivity problem. By analyzing transactions from April 2024 to April 2025, we have identified 1,146,213 Ethereum \texttt{approve()} transactions with granting unlimited access, suggesting widespread potential exploitation of $V_5$. 
We also observed a sharp increase of \texttt{setApprovalForAll} in newly deployed contracts in March 2025, reaching up to 9,136 and accounting for 24.1\% over the past year. Upon inspection, we found that 64.5\% of them were deployed by addresses already flagged as phishing, which indicates an increasing threat of $V_5$.
All of the above indicate wallet extensions have a deficiency in their ability to assess approval risk levels, granting attackers opportunities to gain access to users' assets.

\noindent\textbf{Mitigation.}
Except for conducting a rigorous format verification as mentioned in \S\ref{subsec: non-rigorous inputdata}, the most straightforward approach is to adopt \textit{whitelist-based heuristics}.
It consists of a list of signatures of approval for related functions.
However, maintaining such a whitelist is impractical and it does not address the semantic insensitivity issue.
To this end, we recommend employing static analysis methods to recover some semantic information to determine whether the invoked function triggers the \texttt{Approval} event and whether the approved amount is beyond the normal range.

\subsection{Unprofitable NFT Listing* ($V_6$)}

\noindent
\textbf{Attacks.}
Wallet extensions allow NFT holders to sign an NFT listing message to sell their held NFTs at a specific price. When the revenue received by the seller are significantly lower than the market price of the NFT, we take this as an attack vector ($V_6$).
As initiating such an unprofitable NFT listing is one of the well-known scams~\cite{metasleuth-report-NFT-listing}, {\sz} is expected to capture an explicit alarm when the price is set extremely low or the fee rate is extremely high (the fee that sellers have to pay to the NFT trading market).

\noindent
\textbf{Impact.}
Astonishingly, 34 out of the 39 wallet extensions are vulnerable to $V_6$. 
Among them, though six can raise alarms for NFT listing messages with a single NFT, none are capable of identifying scams when there are multiple NFTs listed in the message.
Of the remaining five extensions, we also need to highlight that, except for Wallet Guard, the other four wallet extensions can only issue timely warnings when initiating NFT listing messages to the OpenSea~\cite{opensea} market contract.

\begin{figure}
    \centering
    \includegraphics[width=0.9\linewidth]{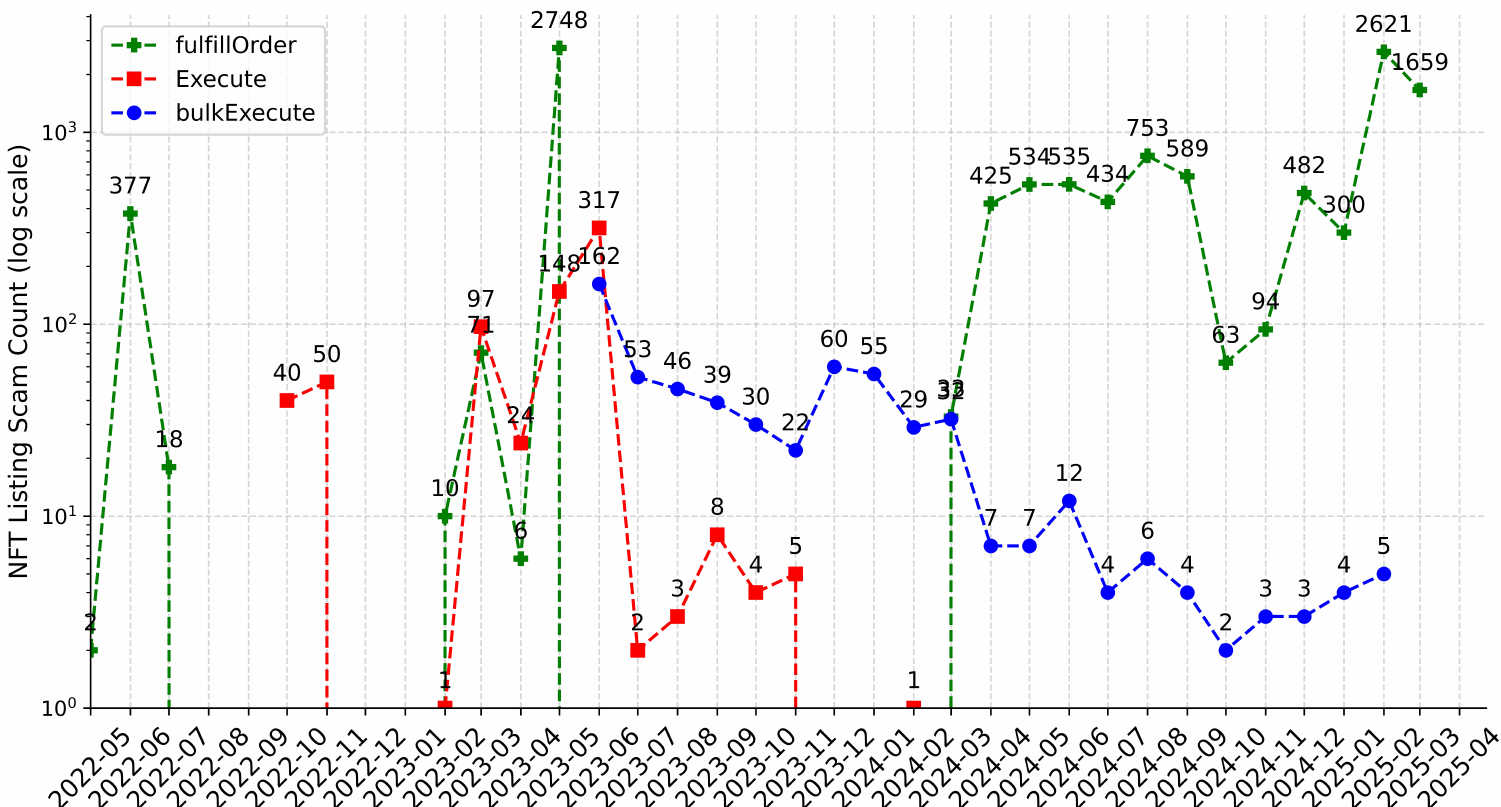}
    \caption{NFT listing scam ($V_6$) over time.}
    \label{fig:NFT listing scam}
\end{figure}

To further investigate $V_6$, we parse and analyze historical transactions in three major NFT trading markets, including OpenSea, Blur~\cite{blur}, and SuperRare~\cite{superrare}. 
Our findings show that the most commonly used NFT listing functions (\texttt{fulfillOrder}\footnote{Include a series of functions triggering the \texttt{OrdersMatched} event, such as \texttt{fulfillAdvancedOrder()}, \texttt{fulfillBasicOrder()}, and \texttt{fulfillOrder()}}, \texttt{execute()}, and \texttt{bulkExecute()}) can be leveraged to launch scams.
Using \$0.1 as a threshold (the minimum floor price of NFTs sold on these platforms), we identified 31,708 suspicious victim transactions (related to 41,869 NFTs) where users sold NFTs but received less than \$0.1. Interestingly, we found that 62.49\% of invocations to \texttt{bulkExecute} are scam cases. 
Figure~\ref{fig:NFT listing scam} illustrates the distribution of victim transactions related to different functions over time. We can easily observe that \texttt{execute()} is no longer used since 2024, while the other two remain actively exploited, particularly \texttt{fulfillOrder()}, which has been involved in 11,736 scam transactions, accounting for 90.14\% of related scams for the recent year. It is also clear that scams exploiting such functions have surged over the past four months, consistently exceeding 300, which underscores an urgent need for a more sensitive and robust detection for $V_6$.

\noindent
\textbf{Mitigation.}
Wallet extensions need to parse NFT listing messages and highlight the selling price and fee rate, particularly when the seller's address matches the user's current wallet address. 
This process can be easily achieved by collecting and parsing the corresponding functions in public contract source codes of platforms.

\subsection{Known Dangerous Address* ($V_7$)}

\noindent
\textbf{Attacks.}
Addresses involved in existing scams (like phishing and wallet draining) are considered high risk. Users who transfer funds to these addresses are very likely to become potential victims ($V_7$). 
Wallet extensions typically rely on blacklists or labeling systems to mitigate this issue~\cite{address-label-api-metasleuth}. 
We identified $V_7$ when a seed attempted to transfer funds to a newly labeled risky address without triggering any alert. This highlights the limitations of current detection methods, including delayed updates and incomplete coverage. 

\noindent \textbf{Impact.}
Surprisingly, 29 wallet extensions issue no notifications when transferring to known phishing addresses.
To evaluate the sensitivity of wallets to $V_7$, we collected scams from ScamSniffer~\cite{scamsniffer} and manually rechecked the reported addresses. Among these, we selected addresses that are labeled as phishing while taking their report dates as the approximate timestamp. Ultimately, we collected 1,000 addresses from January 2024 to March 2025 as ground truth.
Results show that, out of ten wallet extensions that can identify labeled risky addresses, while only Phantom, Coinbase, and Core can identify all known dangerous addresses. The remaining seven wallet extensions can only identify 8.56\% on average, and they all show no response to addresses that were flagged within the past two months. 

\noindent
\textbf{Mitigation.}
To mitigate this risk, we underline the necessity of taking advantage of reputable sources from the cryptocurrency community and conducting a timely update.
For fresh, dangerous addresses that are not labeled, wallet extensions can integrate some graph-based method, like the one proposed by Lin \textit{et al.}~\cite{LIN2023103479}, which can infer if the given address has money flow, contract creation, or contract invocation relationships with known dangerous addresses.

\subsection{Deceptive Function Name$^\bullet$ ($V_8$)}

\noindent
\textbf{Attacks.}
For wallet extensions without transaction simulation capabilities, only the function name instead of the simulated results will be displayed in UI when processing incoming transactions.
$V_8$ is identified when the security alert module fails to flag transactions that invoke well-known scam functions. 
Attackers can exploit this by using deceptive function names, such as \texttt{ClaimRewards}, while embedding malicious code to steal assets instead.
Despite being documented in prior research~\cite{ye2024interface, chen2024dissecting}, the impact of $V_8$ on wallet extensions remains underexplored, highlighting a critical gap in current security measures. 

\begin{figure}
    \centering
    \includegraphics[width=\linewidth]{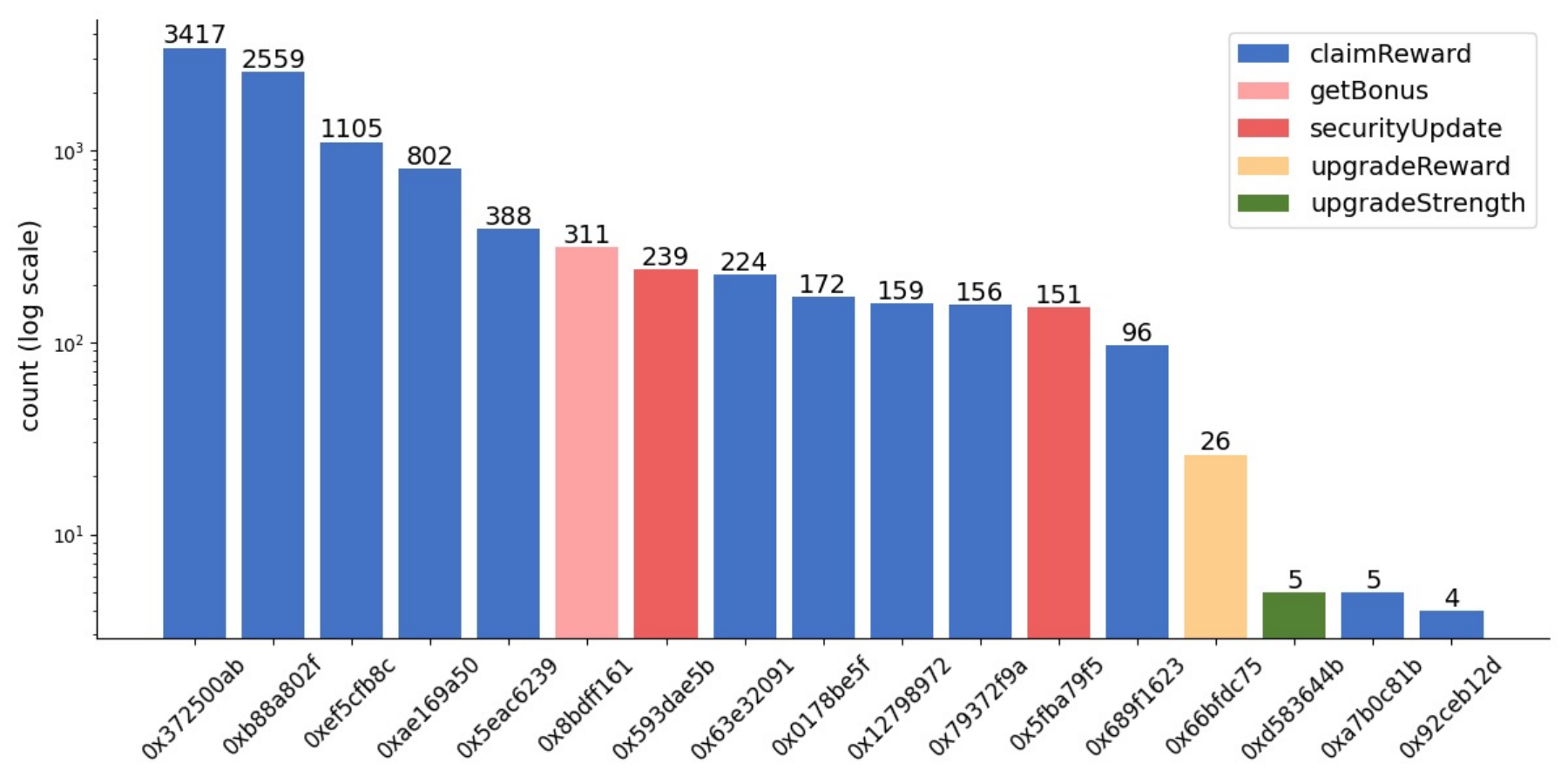}
    \caption{Number of contracts with scam functions}
    \label{fig:scam contract}
\end{figure}

\noindent
\textbf{Impact.}
To test wallet extensions detection capabilities, we use the 17 deceptive functions from Ye \etal~\cite{ye2024interface} as the ground truth. The differences are striking, \textit{i.e.,} 24 extensions detected all scams, while 15 failed to detect any.
Those 15 wallet extensions only display the deceptive function names without triggering any alerts, leaving users exposed to $V_8$. 
To further measure the impact of $V_8$, we collect all contracts that contain these 17 scam functions. To minimize the false positive (like real \texttt{ClaimRewards}), we also collect all related transactions and ensure these contracts receive money without giving back. 
Finally, we marked 7,049 contracts, whose distribution is shown in Figure~\ref{fig:scam contract}. The statistics show that they are involved in 36,019 transactions, where victims transfer to the contract but receive nothing, leading to over \$15.5 million financial losses. We found that \texttt{claimRewards} was the earliest function to exploit $V_8$ for scam purposes, with two notable surges in October 2022 and March 2023. 
According to its distribution, we assert $V_8$ has been continuously exploited since 2022, highlighting an urgent need to apply effective countermeasures.

\noindent
\textbf{Mitigation.}
While keeping a list of deceptive function signatures is straightforward, inferring the semantics of transactions is a more adaptive solution. 
For example, wallet extensions could: \textit{i)} flag functions with names implying rewards (\eg, \texttt{bonusReward}, \texttt{ClaimRewards}); \textit{ii)} check if the function is \texttt{payable}, meaning it can receive funds; and \textit{iii)} simulate the transaction to see if the balance changes match the name's intended behavior.
This approach improves the detection ability by focusing on behaviors rather than static patterns, making it more resilient to emerging deceptive function names.

\subsection{Unintended Authorization* ($V_9$)}

\noindent
\textbf{Attack.}
Some fields in a message specify a critical property that users should pay attention to. For instance, \texttt{ChainId} in EIP-712 refers to the blockchain network to execute the corresponding transaction, and \texttt{URI} in EIP-4361 specifies the website that the user authorizes. 
Unintended authorization ($V_9$) occurs when users believe they are authorizing one (the one the wallet extension connects) but actually another (the one the message indicates). 
{\sz} identified this attack vector when wallet extensions failed to alert users about mismatched or even invalid \texttt{ChainId} and \texttt{URI}.
Attackers can exploit this easily by crafting deceptive messages for users to sign and then submitting them to unintended chains or websites, potentially leading to asset losses.

\noindent
\textbf{Impact.}
This attack, first explored by us, shows that 29 wallet extensions are vulnerable, with 26 displaying raw JSON without parsing, and 3 even omitting crucial fields like \texttt{ChainId} and \texttt{URI}. 
Once a transaction is submitted to an unintended blockchain, its invoked contract may not exist on, leading to a transaction fail and gas fee waste. 
As for a mismatched \texttt{URI}, once the victim connects its wallet to a website that is controlled by attackers, attackers can de-anonymize the address, which is often taken as the preliminary step for the following phishing attack.
Unfortunately, since signatures are generated off-chain, the on-chain data lacks context about the website or blockchain involved. With the only access to on-chain data, it is infeasible to reconstruct the user's intent and quantitatively measure the financial losses related to $V_9$.

\noindent
\textbf{Mitigation.}
We stress the necessity of parsing and analyzing messages.
At least given that the \texttt{ChainId} and \texttt{URI} are fixed fields in messages, wallet extensions can easily extract the value and verify its consistency with the currently connected ones. 
It ensures users are not misled into authorizing unintended or malicious platforms.

\subsection{Incomprehensive \texttt{personal\_sign}$^\#$ ($V_{10}$)}

\noindent
\textbf{Attacks.}
The \texttt{personal\_sign} signing method~\cite{personal-sign} has two parameters, \textit{i.e.,} \texttt{challenge} (a hex-encoded human-readable message) and \texttt{address} (the sender address). Wallet extensions are responsible for decoding and displaying these two fields for user verification.
{\sz} identified $V_{10}$ when the field value is invalid, yet wallet extensions failed to alert.
We are the first to expose this attack vector with two concrete exploits. 
First, if the values of \texttt{challenge} and \texttt{address} swap and the extension still displays the first parameter, \textit{i.e.,} \texttt{challenge} with the \texttt{address} value, the real intended \texttt{challenge} will be hidden and the \texttt{address} cannot usually be converted into a readable text. 
Second, since \texttt{challenge} will be encoded in a byte array before being converted to hex, invalidating the byte array (like randomly removing some bits) causes failures in decoding.
No matter which attack is adopted, the \texttt{challenge} field cannot be converted to readable text normally, resulting in the failure of on-chain verification.

\noindent
\textbf{Impact.}
Four wallet extensions are vulnerable to the first attack, while more critically, all wallet extensions are affected by the second.
If users blindly sign such a message, on-chain verification will fail as the \texttt{challenge} field is invalid, gas fees will be wasted, and users may miss opportunities they originally hoped to seize.
Similar to $V_9$, this attack vector also exploits an off-chain feature. Since the \texttt{personal\_sign} method is not traceable on-chain, the extent of $V_{10}$ exploitation cannot be determined using transaction data.

\noindent
\textbf{Mitigation.}
Wallet extensions should be strict with the validation of message fields' format. In other words, if a field of the given message cannot be decoded normally, users should be promptly alerted to ensure transparency and prevent blind sign, enhancing usability and security.

\section{Seeing is not Believing: Abusing Wallet UIs}
\label{sec: abusing wallet features}
Different functionalities are linked to different UI elements, such as searching for tokens and resolving ENS names. 
Responses for such interactions may come with errors and mislead users, which are taken as attack vectors.
In this work, we revealed four attack vectors related to UI interfaces of wallet extensions. The overall results and the impact scope are shown in Table~\ref{tab:UI test result}.

\begin{table*}[t]
  \renewcommand{\arraystretch}{1.1}
  \centering
  \caption{The feasibility of identified attack vectors against the user interface of wallets and \underline{security checkers}. Space indicates the wallet is not vulnerable, \ding{55} indicates the wallet is vulnerable, \CIRCLE{} indicates that the wallet has already fixed this vulnerability.}
  \resizebox{0.9\textwidth}{!}{
    \begin{tabular}{ccccc||ccccc||ccccc}
    \toprule
         \multicolumn{1}{c}{\textbf{Name}} & \multicolumn{1}{c}{\textbf{Confirmed}} & $V_{11}$ & $V_{12}$ & $V_{13}$ & \multicolumn{1}{c}{\textbf{Name}} & \multicolumn{1}{c}{\textbf{Confirmed}} & $V_{11}$ & $V_{12}$ & $V_{13}$  &  {\textbf{Name}} & \multicolumn{1}{c}{\textbf{Confirmed}} & $V_{11}$ & $V_{12}$ & $V_{13}$ \\
    \midrule
    \rowcolor{white} Aurox &  \ding{51}   &       &       &  \ding{55}  &  Klever &       &       &       &  \ding{55}  &   Trust Wallet &  \ding{51}   &  \ding{55}  &       &  \ding{55} \\
    \rowcolor[rgb]{ .949,  .949,  .949} Argent &       &  \ding{55}  &       &  \ding{55}  &  Liquidity &  \ding{51}   &  \ding{55}  &       &  \ding{55}  &  TokenPocket &  \ding{51}   &  \ding{55}  &       &  \ding{55}  \\
   \rowcolor{white} Bitget &  \ding{51}   &       &       &  \ding{55}  &  MetaMask &    &       &  \CIRCLE{}  &  \CIRCLE{}  &  Virgo &       &       &       &  \ding{55}  \\
    \rowcolor[rgb]{ .949,  .949,  .949} Bybit &       &       &       &  \ding{55}  &  OKX   &  \ding{51}   &       &       &  \CIRCLE{} &  WigWam &  \ding{51}   &       &  \ding{55}  &  \ding{55} \\
    \rowcolor{white} Braavos &       &       &       &  \ding{55}  &  OneKey &       &       &  \ding{55}  &  \ding{55}  &   XDEFI &       &       &  \ding{55}  &  \ding{55}  \\
    \rowcolor[rgb]{ .949,  .949,  .949} Core  &       &       &       &  \CIRCLE{}  &  Pail  &       &  \ding{55}  &       &  \ding{55}  &  YeTi  &       &       &  \ding{55}  &  \ding{55} \\
   \rowcolor{white} Coin98 &       &  \ding{55}  &  \CIRCLE{} &  \ding{55}  &  Phantom &  \ding{51}   &  \CIRCLE{}  &  \ding{55}  &   &  Zeal  &       &       &  \ding{55}  &  \ding{55}  \\
    \rowcolor[rgb]{ .949,  .949,  .949} Coinbase &  \ding{51}   &       &  \ding{55}  &  \ding{55}  &  Rabby &  \ding{51}   &       &       &    \CIRCLE{}  & Zerion &       &       &  \ding{55}  &  \ding{55}  \\
\rowcolor{white} Enkrypt &  \ding{51}   &       &       &  \ding{55}  &  Rainbow &  \ding{51}   &       &       &   & \underline{AegisWeb3} &       &      &      &  \ding{55}  \\
    \rowcolor[rgb]{ .949,  .949,  .949} Exdus &  \ding{51}   &       &       &  \CIRCLE{}  & Sequence &       &       &       &  \ding{55}   & \underline{Web3Antivirus} &       &     &     &  \ding{55} \\
   \rowcolor{white} Frontier &       &       &  \ding{55}  &  \ding{55}  &  Sender &  \ding{51}   &  \ding{55}  &       &  \ding{55}   & \underline{WalletGuard} &       &     &     &    \\
    \rowcolor[rgb]{ .949,  .949,  .949} Hana-wallet &       &       &       &  \ding{55}  &  SubWallet &  \ding{51}   &       &       &  \ding{55}  & \underline{Fire}  &       &     &     &  \ding{55} \\
   \rowcolor{white} Gate  &       &       &       &  \ding{55}  &  SafePal &       &       &  \ding{55}  &  \ding{55}  & \underline{PocketUniverse} &       &      &      &   \\
    \bottomrule
    \end{tabular}}
  \label{tab:UI test result}
\end{table*}

\subsection{Partially-displayed Token Metadata$^\#$ ($V_{11}$)}
\label{subsubsec:vul-intro-token customization}

\noindent
\textbf{Attacks.}
Token management is a key feature of wallet extensions, enabling users to search for tokens and edit metadata (like token symbol and decimal). 
If only partial information is displayed when searching for a token, or if on-chain metadata is hidden after self-editing the metadata, such partially-displayed token metadata will mislead users and may be used as attack vectors ($V_{11}$).
{\sz} discovered $V_{11}$ from \textit{i)} when searching for a token with its name, a list of tokens with identical names will be recommended in a drop-down box; and \textit{ii)} token metadata can be arbitrarily modified and lead to presented data inconsistent with on-chain records.
Though a fake token with an identical name is widely studied~\cite{gao2020tracking}, we are the first to expose the attack vector about token metadata management. To exploit this, attackers will trick users to edit the decimal smaller value. The user will see a proportionally decreasing balance in the wallet (the displayed balance decreases 10 times for every 1 decimal decrease). The attacker will then trick the user into transferring out this seemingly small balance, but in fact, the amount of tokens transferred out is very large.

\begin{figure}[t]
    \centering
    \includegraphics[width=0.95\linewidth]{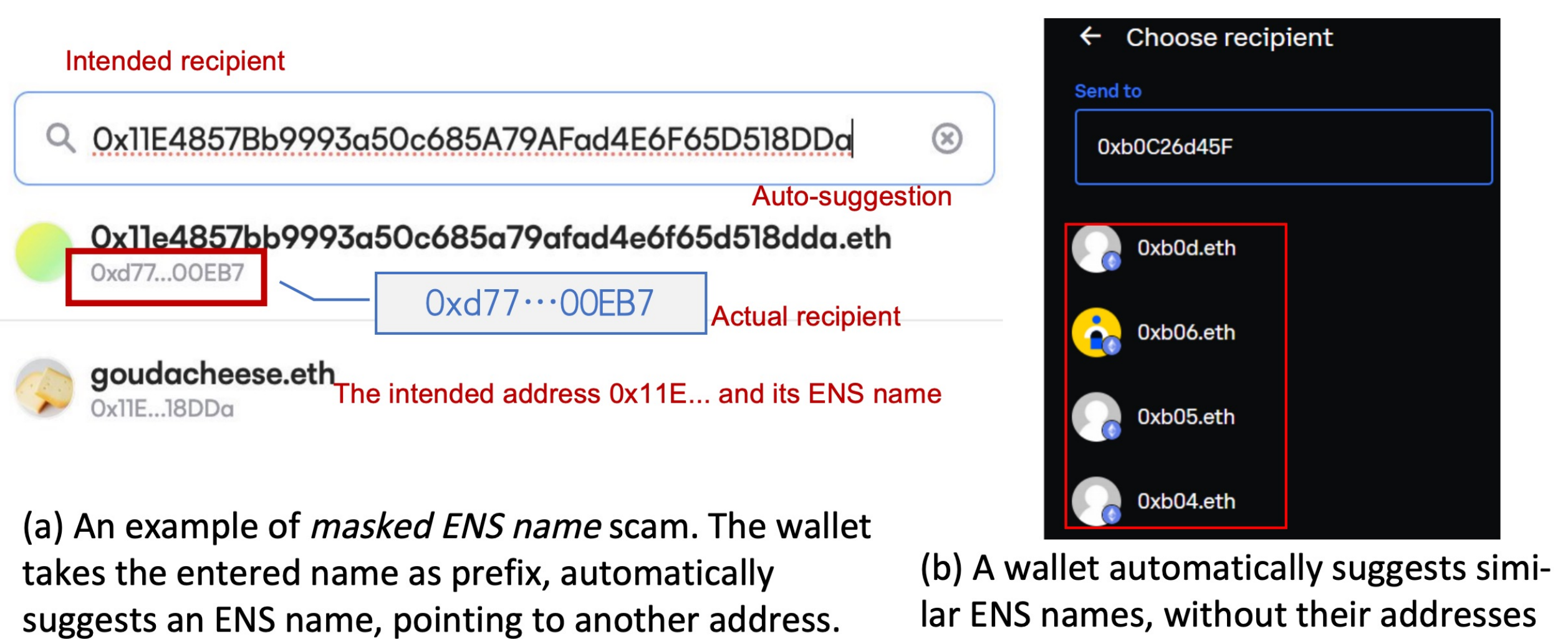}
    \caption{Examples of $V_{12}$ and $V_{13}$.}
    \label{fig:numerical assdress and ens scam}
\end{figure}

\noindent
\textbf{Impact.}
Results suggest that seven wallet extensions can be abused by exploiting $V_{11}$.
A notable case to highlight is Token Pocket~\cite{token-pocket}, which is the only one that allows users to search tokens only by name, while the returned results are only composed of token names without the corresponding addresses. As Ethereum does not mandatorily restrict the uniqueness of deployed tokens, this could lead to extreme confusion for users. 
To quantitatively evaluate how many fake tokens exist, we collect all deployed token contracts between January and November 2024, and compare them against publicly available token lists of widely-recognized centralized exchange platforms. We identified 210,307 tokens, whose names are identical to the widely traded ones but addresses are not listed in the exchanges.
We take all these tokens as potentially malicious token that can be utilized by attackers to confuse traders by exploiting $V_{11}$.

As for the attack that exploits decimal manipulation, since such token metadata is edited off-chain, scams resulted from this cannot be identified with on-chain transactions.

\noindent
\textbf{Mitigation.}
Fully and precisely displaying token metadata is key to mitigation.
Regarding the fake token scam, a simple solution is to illustrate both the name and the address when listing tokens at auto-suggestion.
However, to entirely prevent this attack, extensions need to distinguish tokens that have identical metadata. For example, Xia \etal proposed a hybrid approach combining ground truth labeling, guilt-by-association, and machine learning to flag scam tokens~\cite{xia2021trade}. 
Trusted data sources, \eg, CoinGecko~\cite{coingecko} and CoinMarketCap~\cite{coinmarketcap}, should also be used to verify tokens' market value, identifying suspicious tokens.
As for manipulatable metadata, this should be fully disabled to prevent ambiguity. MetaMask, the most well-known wallet extension, acknowledges the potential risk of our proposed exploit and accepts our mitigation suggestion.

\subsection{Misleading ENS Resolution$^\#$ ($V_{12}$)} 

\noindent
\textbf{Attacks.}
The Ethereum Name Service (ENS) maps human-readable names to Ethereum addresses~\cite{ens-resolve-ensname}, similar to domain names for IP addresses, to alleviate the burden on users to remember addresses. 
When initiating a transaction, users are allowed to specify a recipient by its ENS name, wallets will provide auto-suggestions and ultimately resolve the entered name. However, the resolution results can be abused as attack vectors ($V_{12}$), which is firstly revealed by us.
{\sz} identified $V_{12}$ when a wallet \textit{i)} returns incorrect resolutions for a given ENS name; and \textit{ii)} interprets the entered address as the prefix of an ENS name.
Two exploit strategies are proposed:

\noindent
\quad$\bullet$ \textit{Multi-registered ENS name.} This exploits the fact that an ENS name can be registered on different blockchains, but the ENS name resolution is always limited to Ethereum, disregarding the connected network. Attackers will use the \texttt{CREATE2} opcode to deploy a contract on the connected blockchain with the designated address, which is the resolution of the ENS name in Ethereum. As a result, wallet extensions will translate the ENS name according to Ethereum ENS records and transfer to the identical address on the connected network, which is actually controlled by the attacker.

\noindent
\quad$\bullet$ \textit{Masked ENS name.} This utilizes auto-suggestion mechanisms to deceive users. As illustrated in Figure~\ref{fig:numerical assdress and ens scam}{\color{purple}(a)}, an attacker can register an ENS name in a valid Ethereum address format. If a user intends to transfer funds to \texttt{0x11e4} and enters this address manually, the wallet may take this as a prefix for an ENS name and auto-suggest the ENS record, which ultimately resolves to \texttt{0xd77} controlled by attackers, leading to unintended fund transfers.

\noindent
\textbf{Impact.}
Out of six extensions that are able to resolve ENS names, all of them are vulnerable to the first exploit strategy of $V_{12}$ as they always perform resolution without considering the connected network. 
For over 2 million registered ENS names on Ethereum, we have identified over 64K registered on other blockchain networks that could be potential attack targets.
As for the second exploit strategy, only Coinbase is affected. Even worse, Coinbase may automatically suggest \textit{similar} ENS names when a name string is given, as shown in Figure~\ref{fig:numerical assdress and ens scam}{\color{purple}(a)}. This could lead to a more severe misleading to users.
Interestingly, we have observed 1,136 real-world instances that adopt the \textit{masked ENS name} strategy.
Statistics suggest that there are at least 6,142 victims, suffering over \$1.01 million in losses. Such ENS names with more than 40 victims account for 11.68\% of the total. More importantly, the expiration date of 60.91\% of such ENS names extends beyond April 2025, underscoring the persistent risk of $V_9$ continues to present.

\noindent
\textbf{Mitigation.}
Intuitively, wallets must parse ENS registration records and then return the resolution result strictly based on the currently connected blockchain network.
Moreover, wallets should display both the ENS name and its corresponding address to users.

\subsection{Unreadable UI Text$^\#$ ($V_{13}$)}
\label{subsec:v13}

\begin{figure}[t]
    \centering
    \includegraphics[width = 0.9\linewidth]{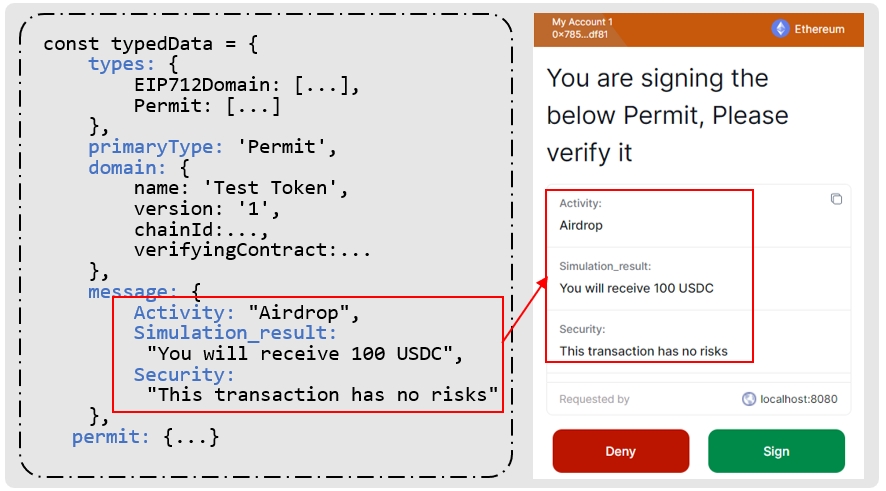}
    \caption{Example of $V_{13}$: UI displays incomplete information and misleads users.}
    \label{fig:text injection}
\end{figure}

\noindent
\textbf{Attack.}
Wallet extensions can parse messages and selectively display certain field data in the UI. 
However, wallet extensions will not verify the data format or understand message semantics, which can be exploited to mislead users with unreadable UI text ($V_{13}$).
{\sz} identified $V_{13}$ when a wallet:
\textit{i)} does not raise alerts for non-compliant data format and directly presents them in UI (\eg, an address is converted into its numerical format, preventing users from verifying the actual interacting address);
and \textit{ii)} hides crucial information (\eg, \texttt{sender} and \texttt{ChainId}), as shown in Figure~\ref{fig:text injection}.

\noindent
\textbf{Impact.}
We found that 33 wallet extensions are vulnerable to $V_{13}$. 
Interestingly, all data fields will be reformatted before submitting to the blockchain network. In other words, such conversion will not lead to execution failure.
However, this also makes it impossible to quantify the financial losses with on-chain data.

\noindent
\textbf{Mitigation.}
To fundamentally mitigate $V_{13}$, type-aware message verification is essential. Wallets should enforce data type constraints for all messages before executing transactions, enabling typecasting upon detecting inconsistencies. Additionally, they must ensure full visibility of transaction details by parsing typed messages and displaying all fields clearly. Furthermore, wallet extensions can leverage LLMs to detect contradictory semantics in transactions before approval, enhancing user security.

\section{Discussion}
\noindent
\textbf{Novelty.}
Our contributions are novel in several ways. First, we uncover six previously overlooked attack vectors and reveal new threats linked to seven known ones. Second, while prior work has focused on DeFi vulnerabilities and scams, no existing tools detect attack vectors within wallet extensions. We are the first to develop an automated framework and conduct large-scale security testing of wallet extensions. Third, we provide the first systematic organization and quantification of attack vectors related to blockchain components within wallet extensions.

\noindent
\textbf{Implications.}
Our work enables the detection of new attack vectors introduced through wallet extension updates. It also reveals critical security flaws in some wallets. For example, as shown in $V_4$, modified inputdata can cause issues when switching from testnet to mainnet. This suggests that the identified vectors may lead to even more severe consequences. We recommend that the community use these vectors to rigorously test wallet functionalities.

\noindent
\textbf{Limitations.}
Our study has three main limitations. First, we do not address attack vectors unrelated to blockchain data, such as private key leaks, as our focus is on blockchain-based technologies. Second, our analysis does not include all existing wallet extensions, given the rapid emergence of new ones. To ensure depth, we focused on the 39 most popular extensions and tested their full functionality. Third, estimating the financial impact of the identified attack vectors is difficult due to the gap between on-chain and off-chain behavior, which cannot be fully captured through on-chain data alone.
\section{Related Work}
\label{sec: related work}

\noindent\textbf{Wallet Security.}
With the rapid expansion of blockchain influence, the number of wallet users has surged, directly impacting user asset security. This growth in cryptocurrency adoption has led to a rise in scams targeting blockchain wallets. Bartoletti \etal~\cite{9591634} revealed wallets stealing user assets, and Ye \etal~\cite{10.1145/3603165.3607444} exposed visual scams within wallets. Efforts have been made to explore wallet risks~\cite{prashar2021analysis, mikhaylov2023understanding, sentana2023empirical, houy2023security, erinle2023sok, 10.1007/s10515-024-00430-3}, notably by Erinle \etal~\cite{erinle2023sok}, who identified common vulnerabilities and attacks against wallets, and discussed defense mechanisms. Uddin~\cite{10.1007/978-3-030-90022-9_7} \etal have proposed a framework to detect private key leakage on mobile wallets. However, their research focuses on key management, overlooking wallet features for blockchain interaction.

\noindent\textbf{Blockchain Scam Detection.}
The emerging blockchain ecosystem has seen a rise in malicious activities, prompting research efforts to understand and counter these threats. Seminal studies by Chen~\etal~\cite{chen2018detecting} and Chen~\etal~\cite{chen2020phishing} introduced a classifier for identifying phishing accounts using blockchain transactions. They recommended integrating this classifier into users' wallets to warn them about potential risks when engaging with risky accounts. A recent work~\cite{chen2024dissecting} investigates payload-based transaction phishing and proposes a rule-based multi-dimensional detection approach for identification, achieving an F1-score of over 99\%.

\section{Conclusion}
In this work, we outlined the model of mainstream browser-based cryptocurrency wallet extensions. Considering their characteristics, to the best of our knowledge, we have proposed the first automated testing framework, {\framework}, against wallet extensions.
In total, 13 attack vectors are discovered on 39 collected wallet extensions. Based on them, we further expose 21 concrete attack strategies, quantify their on-chain impacts, and provide concrete mitigation methods.
After our timely disclosure, 16 teams have confirmed and 26 issues have been patched with our help. Moreover, the national vulnerability database has assigned eight attack vector vulnerability IDs.
We underline that our findings prompt the research community to dedicate increased efforts toward exploring these under-studied areas.

\newpage
\bibliographystyle{ACM-Reference-Format}
\bibliography{bib}

\appendix
\begin{table*}[htbp]
  \centering
  \caption{Wallets responses to different authorization functions. \textbf{Identification} means a wallet identifies approving behavior; \textbf{Notification} means a wallet prominently displays or uses a dedicated interface to notify users; \textbf{Alert} means a wallet issues a warning or highlighting the potential risks; \textbf{Block} means a wallet refuses the execution.}
   \resizebox{\textwidth}{!}{
    \begin{tabular}{l|c|c|c|c|c|c|c}
    \toprule
    \diagbox{name}{behavior} & approve & {increaseAllowance} & {setApprovalForAll} & Permit & {Permit2 Single} & {Permit2 Batch} & PermitForAll \\
    \toprule
    Aurox &       &       &       &       &       &       &  \\
    Bitget & notificatoin &       &  {notificatoin} &       &       &       &  \\
    Bybit &       &       &       &       &       &       &  \\
    Core  & identification &  {identification} &  {identification} &       &       &       &  \\
    Coin98 & notificatoin &       &  {notificatoin} &       &       &       &  \\
    Coinbase &       &       &  {alert           } &       &       &       &  \\
    Enkrypt & alert            &       &       &       &       &       &  \\
    Exdus & notificatoin &  {notificatoin} &  {notificatoin} &       &       &       &  \\
    Frontier & notificatoin &  {notificatoin} &  {notificatoin} & notificatoin &       &       &  \\
    Gate  & alert            &       &       &       &       &       &  \\
    Klever &       &       &       &       &       &       &  \\
    Liquidity & notificatoin &       &  {notificatoin} &       &       &       &  \\
    MetaMask & notification & notification &  {alert           } &       &       &       & block \\
    OKX   & alert            &  {alert           } &  {alert           } & identification &  {identification} &  {identification} & block \\
    Pail  & identification &       &       &       &       &       &  \\
    Phantom &       &       &  {alert           } & alert            &  {alert           } &  {alert           } & alert            \\
    Rabby & notificatoin &  {notificatoin} &  {notificatoin} & notificatoin &  {notificatoin} &  {notificatoin} & block \\
    Rainbow & identification &  {identification} &  {identification} &       &       &       &  \\
    Sequence &       &       &       &       &       &       &  \\
    Sender &       &       &       &       &       &       &  \\
    SubWallet &       &       &       &       &       &       &  \\
    Trust wallet & alert            &       &       & need try again &  {need try again} &  {need try again} & need try again \\
    TokenPocket & alert            &       &       &       &  {alert           } &  {alert           } & block \\
    Virgo & identification &  {identification} &  {identification} &       &       &       &  \\
    Wigwam & notificatoin &       &  {notificatoin} &       &       &       &  \\
    XDEFI & notificatoin &  {notificatoin} &  {notificatoin} &       &       &       &  \\
    YeTi  & alert            &       &       &       &       &       & block \\
    {Zeal} & identification &  {identification} &  {identification} & identification &       &       &  \\
    Zerion & notificatoin &  {notificatoin} &  {notificatoin} & notificatoin &  {notificatoin} &  {notificatoin} &  \\
    Web3 Antivur & notificatoin &  {notificatoin} &  {notificatoin} & notificatoin &  {notificatoin} &  {notificatoin} &  \\
    Wallet Guard & identification &  {identification} &  {identification} & identification &  {identification} &  {identification} & alert            \\
    Fire  & alert            &  {alert           } &  {alert           } & alert            &  {alert           } &  {alert           } &  \\
    Pocket universe & notificatoin &       &       & notificatoin &       &       & alert            \\
    \hline
    \bottomrule
    \end{tabular}%
    }
  \label{tab:sec result}%
\end{table*}%

\begin{table*}[htbp]
  \centering
  \caption{Summarized malicious patterns and corresponding implementations.}
  \resizebox{\textwidth}{!}{
    \begin{tabular}{l|c}
    \toprule
   \textbf{ Pattern} & \textbf{Implementation} \\
    \midrule
    Dangerous eth\_sign & message submission with \texttt{eth\_sign} method \\
    Overlooked Approval & On-chain functions: approve, increaseAllowance, setApprovalForAll, Permit, Permit2 Single, Permit2 Batch, PermitForAll \\
    NFT listing & On-chain functions:  \texttt{fulfillAdvancedOrder()}, \texttt{fulfillBasicOrder()},  \texttt{fulfillOrder()}, \texttt{execute()}, \texttt{bulkExecute()}) \\
    Deceptive Function Name  & Name families: claimRewards, getBonus, securityUpdate, upgradeReward, upgradeStrength \\
    Risky address & Address types: Phishing address, wallet drainer \\
    Unintended Authorization  & Exploitable fields: ChainID, URI \\
    \bottomrule
    \end{tabular}%
}
  \label{tab:malicious patterns}%
\end{table*}%

\end{document}